\documentclass[sigconf,nonacm]{acmart}

\usepackage{microtype}
\usepackage{multirow}
\usepackage[normalem]{ulem}
\usepackage{graphicx}
\usepackage{blindtext}
\usepackage{subfiles}
\usepackage{subcaption}
\usepackage{float}
\usepackage{afterpage}
\usepackage{color}
\usepackage{enumitem}
\usepackage{amsmath}

\useunder{\uline}{\ul}{}
\graphicspath{images/}
\AtBeginDocument{%
  }

\setcopyright{acmlicensed}
\copyrightyear{2025}
\acmYear{2025}
\acmDOI{XXXXXXX.XXXXXXX}

\acmConference[KDD '25]{the 31st ACM SIGKDD Conference on Knowledge Discovery and Data Mining}{August 03--07, 2025}{Toronto, Canada}

\acmISBN{978-1-4503-XXXX-X/18/06}

\title{Embed Progressive Implicit Preference in Unified Space for Deep Collaborative Filtering}
                             
\author{Zhongjin Zhang}
\authornote{Equal contribution.}
\affiliation{%
  \institution{Central South University}
  \city{Changsha}
  \country{China}}
\email{zhangzhongjin@csu.edu.cn}

\author{Yu Liang}
\authornotemark[1]
\affiliation{%
  \institution{Central South University}
  \city{Changsha}
  \country{China}}
\email{yl8373713@gmail.com}

\author{Cong Fu}
\affiliation{%
  \institution{Shopee Pte. Ltd.}
  \city{Singapore}
  \country{Singapore}}
\email{fc731097343@gmail.com}

\author{Yuxuan Zhu}
\affiliation{%
  \institution{Shopee Pte. Ltd.}
  \city{Shanghai}
  \country{China}}
\email{iamyuxuanzhu@gmail.com}

\author{Kun Wang}
\affiliation{%
  \institution{Shopee Pte. Ltd.}
  \city{Shanghai}
  \country{China}}
\email{wk1135256721@gmail.com}

\author{Yabo Ni}
\affiliation{%
  \institution{Nanyang Technological University}
  \city{Singapore}
  \country{Singapore}}
\email{yabo001@e.ntu.edu.sg}

\author{Anxiang Zeng}
\affiliation{%
  \institution{SCSE, Nanyang Technological University}
  \city{Singapore}
  \country{Singapore}}
\email{zeng0118@e.ntu.edu.sg}

\author{Jiazhi Xia}
\authornote{Corresponding author.}
\affiliation{%
  \institution{Central South University}
  \city{Changsha}
  \country{China}}
\email{xiajiazhi@csu.edu.cn}

\newcommand{\rv}[1]{\mathrm{#1}}         % a scalar random variable 正体
\newcommand{\vecmy}[1]{\boldsymbol{#1}}  % a vector 加粗的斜体
\newcommand{\vecrv}[1]{\mathbf{#1}}      % a vector-valued random variable 加粗的正体

\usepackage{tcolorbox}

\renewenvironment{boxed}
    {
    \begin{tcolorbox}[colback=black!2!white,colframe=black!75!black,left=2pt,right=2pt,top=2pt,bottom=2pt]
    }
    { 
    \end{tcolorbox}
    }

\begin{document}

\begin{abstract}
Embedding-based collaborative filtering, often coupled with nearest neighbor search, is widely deployed in large-scale recommender systems for personalized content selection. Modern systems leverage multiple implicit feedback signals (e.g., clicks, add to cart, purchases) to model user preferences comprehensively. However, prevailing approaches adopt a feedback-wise modeling paradigm, which \textbf{(1)} fails to capture the structured progression of user engagement entailed among different feedback and \textbf{(2)} embeds feedback-specific information into disjoint spaces, making representations incommensurable, increasing system complexity, and leading to suboptimal retrieval performance. A promising alternative is Ordinal Logistic Regression (OLR), which explicitly models discrete ordered relations. However, existing OLR-based recommendation models mainly focus on explicit feedback (e.g., movie ratings) and struggle with implicit, correlated feedback, where ordering is vague and non-linear. Moreover, standard OLR lacks flexibility in handling feedback-dependent covariates, resulting in suboptimal performance in real-world systems. To address these limitations, we propose Generalized Neural Ordinal Logistic Regression (GNOLR), which encodes multiple feature-feedback dependencies into a unified, structured embedding space and enforces feedback-specific dependency learning through a nested optimization framework. Thus, GNOLR enhances predictive accuracy, captures the progression of user engagement, and simplifies the retrieval process. We establish a theoretical comparison with existing paradigms, demonstrating how GNOLR avoids disjoint spaces while maintaining effectiveness. Extensive experiments on ten real-world datasets show that GNOLR significantly outperforms state-of-the-art methods in efficiency and adaptability.
\end{abstract}

\iffalse

\fi

% \keywords{Ordinal Logistic Regression, Recommender System, Personalized Ranking, Collaborative Filtering, Multi-Task Learning}
\maketitle

\section{INTRODUCTION}
\label{sec:intro}
\begin{figure}[t]
    \centering
    \includegraphics[width=0.96\linewidth]{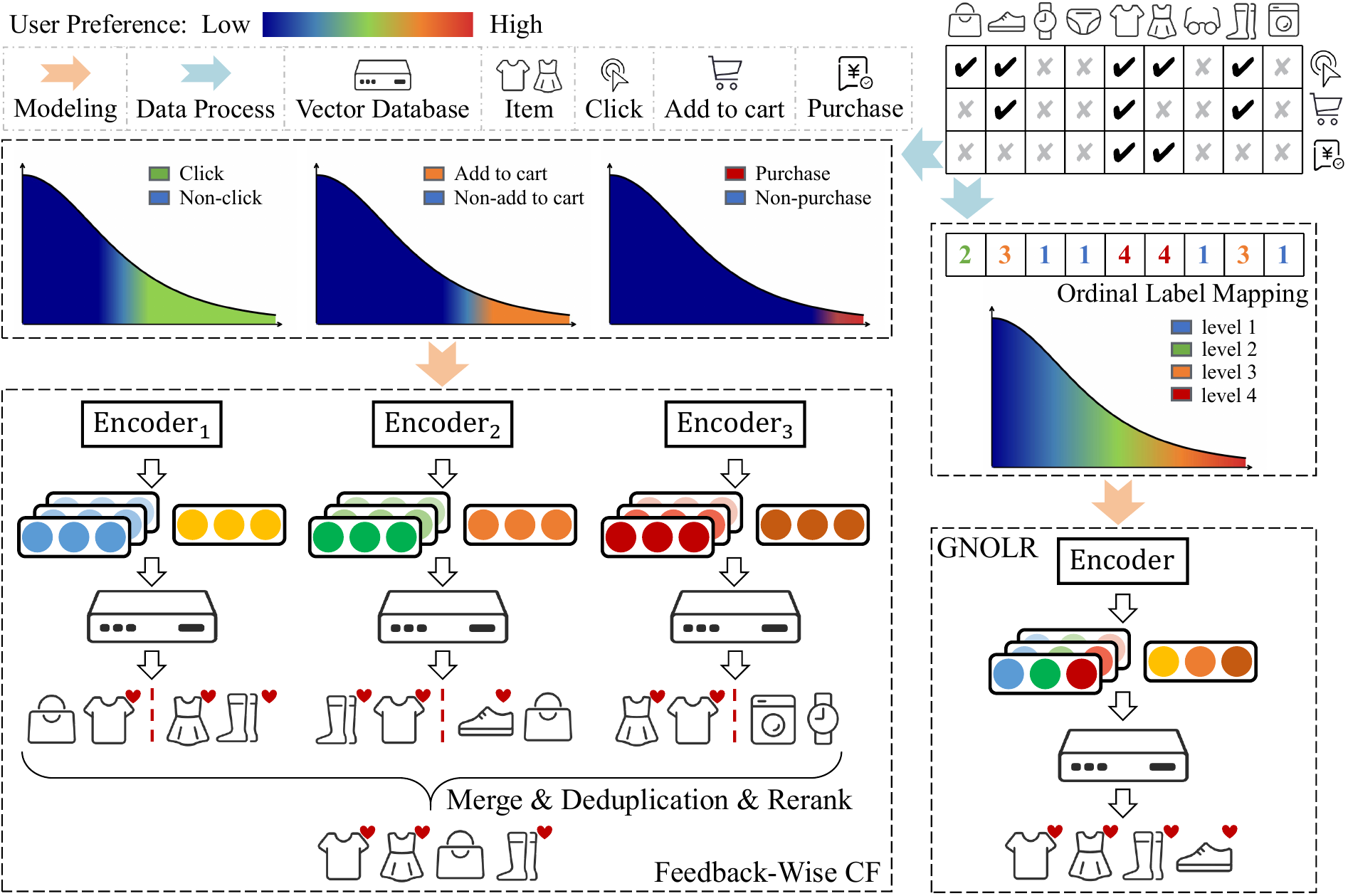}
    \caption{An illustration comparing the prior paradigm with GNOLR. Prior methods embed entities into feedback-wise, incommensurable spaces, requiring separate ranking and fusion before displaying to users. In contrast, GNOLR unifies various feedback in a shared embedding space, aligning the spatial proximity with the user preference progression for seamless one-stage ranking and improved prediction.}
    \Description[GNOLR vs prior methods comparison diagram]{Prior methods embed entities into feedback-wise, incommensurable spaces, requiring separate ranking and fusion before displaying to users. In contrast, GNOLR unifies various feedback in a shared embedding space, aligning the spatial proximity with the user preference progression for seamless one-stage ranking and improved prediction.}
    \label{fig:intropic}
\end{figure}

To mitigate information overload, embedding-based collaborative filtering (CF)~\cite{CFsurvey,NCF} has become the foundation of large-scale recommender systems (RS), enabling personalized item selection and ranking. In practice, CF is often paired with nearest neighbor search (NNS)~\cite{dasgupta2011fast,jegou2010product,malkov2018efficient,fu12fast} for efficient retrieval, ensuring that relevant candidates are quickly surfaced for users~\cite{YoutubeDnn}. Modern RS increasingly rely on multiple implicit feedback signals, e.g., click/like a video or click/purchase a commodity, to infer user preferences~\cite{AITM, DCMT, TAFE, NISE, ResFlow}. These pieces of feedback reflect different levels of user engagement, making it crucial to model them holistically. However, prevalent embedding-based CF methods face \textbf{three fundamental limitations} that hinder their effectiveness: 

\noindent\textbf{(1)} They treat each feedback type as an independent binary classification or ranking task, overlooking the underlying progression in user engagement. For instance, in purchase prediction, clicked but unpurchased items are treated as equally negative as non-clicked items. However, semantically, the former is preferred over the latter. This simplification can lead to misinterpretation of user intent.

\noindent\textbf{(2)} Task-wise modeling produces disjoint embedding spaces for each feedback type, making them incommensurable: the similarity scores computed in one space cannot be meaningfully compared to those from another. This fragmentation prevents a unified candidate retrieval process (Figure~\ref{fig:intropic}). As a result, large-scale RS, which rely on efficient NNS~\cite{airbnb}, suffer from redundant indexing and ranking. In addition, because scores from different tasks follow different scales and distributions, fusion heuristics such as additive or multiplicative aggregation~\cite{MOO,MPE,MTF} often introduce accuracy loss and distort user preferences. This not only increases system complexity as the number of tasks grows but also degrades recommendation quality.

\noindent\textbf{(3)} Task-wise modeling employs separate prediction heads for different feedback labels. When feedback labels contradict each other (e.g., clicked but unpurchased), gradient conflicts arise in shared parameters, negatively impacting learning stability~\cite{yu2020gradient,he2024multibalance}.

Ordinal Logistic Regression (OLR)~\cite{regression1980} has been applied in RS~\cite{OrdRec,hu2018collaborative} to model progressiveness over explicit ratings, offering potential for progressive implicit feedback. However, applying standard OLR in this context also introduces \textbf{key challenges}: 

\noindent\textbf{(1)} Unlike ratings, implicit feedback does not always follow a rigid sequence (e.g., users may follow an influencer without "liking" their videos). This violates OLR's strict ordinal assumption.

\noindent\textbf{(2)} Standard OLR~\cite{regression1980} applies the same set of regression coefficients to all feedback types, assuming a uniform relationship between features and different levels of user engagement. However, feedback like clicks and purchases are influenced by distinct factors (e.g., click-through rate vs. sales count). Simply introducing separate coefficients or networks for each feedback type may address this limitation, but it \textbf{reintroduces disjoint embedding spaces}.

To address these challenges, we propose \textbf{Generalized Neural Ordinal Logistic Regression (GNOLR)}, a novel framework that integrates ordinal modeling with neural representation learning for implicit feedback. Our key contributions are as follows:

\textbf{(1)} We introduce an ordinal mapping technique to transform unstructured implicit feedback into strictly ordered labels, enabling effective ordinal modeling. GNOLR employs a nested optimization framework that captures feedback-specific feature dependencies while embedding underlying user preferences in a unified space, enabling a single NNS process without compromising accuracy.

\textbf{(2)} We provide theoretical analysis demonstrating how GNOLR’s embedding similarity structure aligns with the progression of user engagement and highlight its advantages over traditional feedback-wise multi-task approaches.

\textbf{(3)} Extensive experiments on ten real-world datasets show that GNOLR achieves state-of-the-art performance, significantly improving both ranking accuracy and retrieval effectiveness.
\section{PRELIMINARY}
\subsection{Notations}
We differentiate vectors and scalars in bold font, e.g., a vector $\vecmy{x}$ and a scalar $x$. A random variable and its realization are differentiated by italic font, for example, the vector-valued random variable $\vecrv{x}$ and its realization $\vecmy{x}$. Let $\mathcal{U}$ denote the set of users and $\mathcal{I}$ denote the set of items. Let $Y=\{y_1,y_2,...,y_T\}$ represent the label set corresponding to $T$ types of implicit feedback, where each $y_i \in\{0,1\}$. Let $\mathcal{D} =\{(\vecmy{x_u},\vecmy{x_i}, Y_{u,i})\}$ denote the observed samples, where $\vecmy{x_u},\vecmy{x_i}$ are the feature vectors for the users and the items.

\subsection{Problem Formulation and Challenges}
The ultimate goal of leveraging various implicit feedback signals in RS is to achieve a unified and comprehensive understanding of the preference progression in user interactions. This is often overlooked in current literature. To push the research a step further toward real-world applicability, this paper focuses on a new problem:

\subsubsection{Multi-Feedback Collaborative Filtering (MFCF)} MFCF aims to jointly model multiple types of implicit feedback (e.g., clicks, add to cart, purchases) to generate a \textbf{global ranking of candidate items} for display to users. Formally, for a given user $u\in\mathcal{U}$, the objective is to rank a set of candidate items $\mathcal{I}$ by leveraging all available feedback signals $Y$ to generate a unified global ranking list $\mathcal{L}(u)=[i_1, i_2, ..., i_m]$, where $i_j \preceq_u i_k$ if $j<k$. Here, $\preceq_u$ denotes that user $u$ prefers item $i_j$ over $i_k$. 

In collaborative filtering, this corresponds to learning embeddings for $u$ and $i$, and a scoring function $\mathcal{K}$ such that $\mathcal{K}(u,i_j) > \mathcal{K}(u,i_k)$ implies $i_j \preceq_u i_k$. Predominant paradigms utilize neural networks to extract user and item representations~\cite{DSSM,GoogleRetrieval,DAT} and employ kernel functions for similarity measures~\cite{DSSM,EBR,pEBR}.

\subsection{Ordinal Logistic Regression}
Below is a brief introduction to Ordinal Logistic Regression~\cite{regression1980}, which serves as the foundation of our methodology.
\begin{boxed}
\begin{definition}
\label{def:pom}
(Proportional Odds Model). 
Given the ordinal category $k\in \{0,...,C\}$, the proportional odds model satisfies:
\begin{equation}
\mathop{\log}\frac{P(\rv{k} \leq c \mid\vecmy{x})}{P(\rv{k} > c\mid\vecmy{x})} = a_c - {\beta}^\top \vecmy{x},
\label{eq:pom}
\end{equation}
where \(\rv{k}\) denotes an observed ordinal variable with \( C \) categories. \( \vecmy{x} \) denotes the covariate, and \( \mathbf{\beta} \) represents the corresponding regression coefficients. The thresholds \( a_c \) determine when each level of the ordinal response variable becomes likely. 
\end{definition}
\end{boxed}

For each category \( c \), the cumulative probability is defined as:
\begin{equation}
P(\rv{k} \leq c \mid \vecmy{x}) = \sigma(a_c - {\beta}^\top \vecmy{x})
\label{eq:cdf}
\end{equation}
where $\sigma(\cdot)$ represents the sigmoid function. Especially, $k=0$ is a null category and we define $P(\rv{k} \leq 0) = 0$. Additionally, we have $P(\rv{k} \leq C) = 1$. As a result, the probability of predicting category \( \rv{k} = c \) is defined as:
\begin{equation}
P(\rv{k} = c \mid \vecmy{x}) = P(\rv{k} \leq c \mid \vecmy{x}) - P(\rv{k} \leq c-1 \mid \vecmy{x})
\label{eq:pdf}
\end{equation}

According to the proportional odds assumption, the threshold parameters \( a_c \) are independent, but the regression coefficients \( \mathbf{\beta} \) are shared across all categories. The loss function using Maximum Likelihood Estimation (MLE) to optimize the model is given by:

\smallskip
$
\quad\quad\quad\quad L = - \sum_{s=1}^{S}\sum_{c=1}^{C} I(k_s=c) \log P(\rv{k}_s = c \mid \vecmy{x}_s)
$

\smallskip\noindent
where $I(k_s=c)$ is an indicator function and $S$ is the size of samples. 
\section{METHODOLOGY}
\begin{figure*}[t]
    \centering
    \includegraphics[width=0.94\linewidth]{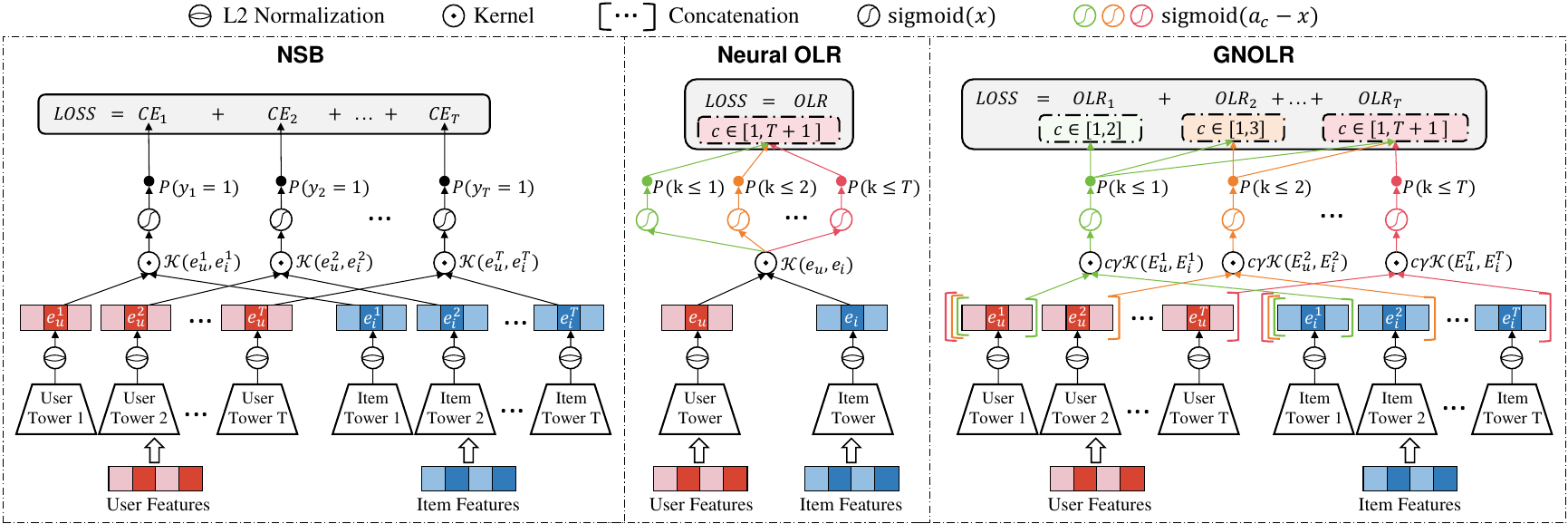}
    \caption{Comparison of three architectures. NSB (left) represents the predominant collaborative filtering framework for multiple implicit feedback signals, known as Naive Shared Bottom, which models each task independently. Neural OLR (middle) extends OLR to neural modeling using a shared encoder. GNOLR (right) further generalizes OLR with nested encoding and subtasks to enhance expressibility and unify the embedding of user engagement across tasks.}
    \Description[Comparison of NSB, Neural OLR, and GNOLR architectures]{This figure compares three model architectures for multi-feedback recommendation. On the left, NSB (Naive Shared Bottom) treats each implicit feedback task independently using a shared base. In the middle, Neural OLR introduces a shared encoder to extend Ordered Logistic Regression into a neural framework. On the right, GNOLR generalizes OLR by introducing nested encoders and subtasks, aiming to improve model expressiveness and unify user engagement representations across tasks.}
    \label{fig:method}
\end{figure*}

To effectively address the MFCF problem and overcome the limitations of prior works, we propose Generalized Neural Ordinal Logistic Regression (GNOLR) framework, comprising two key components: \textbf{(1) a mapping mechanism} that translates unstructured implicit feedback into structured, ordered categorical labels, and \textbf{(2) a generalized OLR model} with enhanced expressibility to capture complex relationships in implicit feedback.

\subsection{Map Implicit Feedback To Ordinal Labels}
To align implicit feedback with the ordinal label constraints of OLR while reflecting users’ progressive engagement, we propose a label mapping mechanism comprising two steps: \textit{(1) Sparsity-based Feedback Ordering} and \textit{(2) Exclusive Ordinal Category Assignment}.

\noindent\textbf{Step 1.} Prior studies~\cite{AITM,ResFlow,TAFE} suggest that feedback sparsity indicates the level of user preference: actions requiring greater commitment (e.g., purchases) occur less frequently but convey higher preference than more common actions (e.g., clicks). We thus arrange feedback signals in ascending order of occurrence frequency, i.e., signals with lower frequency yet higher preference come later.

Formally, consider $T$ feedback types $Y=\{y_t\}_{t=1}^T$, where $y_t=0$ denotes the absence of feedback (often treated as negative~\cite{ESMM,ESCM2,DCMT,TAFE,AITM}). Let $pos(y_t)$ denote the number of samples in which $y_t=1$. We reorder feedback types so that the index $i>j$ if $pos(y_i)<pos(y_j)$. This yields a new sequence $[y_1,\dots,y_T]$ in ascending order of sparsity (i.e., from the lowest preference to highest preference).

\noindent\textbf{Step 2.} Given above ordered feedback list, our goal is to map each sample’s implicit signals to a \emph{single} ordinal label $k$. Because implicit feedback may not follow a strict order and multiple feedback can be positive simultaneously (e.g., user can click and purchase an item without adding to cart), we select the \emph{largest index of positive feedback} within the reordered list, indicating the highest engagement level is reached. Formally, this mapping is defined as:

$
\quad\quad k = 
\begin{cases}
\max(\{t+1 \mid y_t>0\}), & \text{if }\exists y_t > 0, t\in[1,T] \\
1, & \text{if } \forall t\in [1,T], y_t = 0.
\end{cases}
$

\noindent Note that, by default, $P(\rv{k} \le 0)=0$ for $k=0$, to maintain the mathematical rigor of the formulation. $k=1$ denotes the "no positive feedback" (impression) case. Overall, multiple feedback is mapped to one $k\in\{0,T+1\}$, implying the ordinal user preference level.

\textbf{Illustrative Example.} Consider three implicit feedback signals: click ($y_1$), add to cart ($y_2$), and purchase ($y_3$). These events are naturally ordered by their sparsity as $[y_1,y_2,y_3]$. \textbf{Case 1:} If all three labels for an item $i$ are 0 for a user $u$, it means $u$ has only seen the item (impression) without taking any action. This state indicates the lowest engagement, and the mapped label is  $k=1$. \textbf{Case 2:} If $u$ clicked on $i$ and directly purchased it, we have $y_1=1$ and $y_2=0$, and $y_3=1$. Based on the mapping, $k=4$, as purchase ($y_3=1$) represents the highest level of engagement and sparsity.

\subsection{Generalize OLR with Nested Optimization}
With the mapped ordinal labels, we generalize the standard OLR in two aspects to enhance its expressibility. First, we replace its linear formulation with Neural OLR~\cite{cheng2008neural} to model non-linear dependencies between covariants and labels. Second, we introduce a nested optimization framework to capture the progressive structure of user-item interactions (see Figure~\ref{fig:method}).

\subsubsection{Neural OLR for MFCF} To learn user- and item-specific embeddings for the MFCF problem, we adopt the Twin Tower architecture~\cite{DSSM,GoogleRetrieval,DAT}, in which separate neural encoders extract embeddings for users and items. Formally, let $f_u(\cdot)$ and $f_i(\cdot)$ denote neural encoders that map user and item features to their respective embeddings: $\vecmy{e}_u=f_u(\vecmy{x}_u)$ and $\vecmy{e}_i=f_i(\vecmy{x}_i)$, where $\vecmy{x}_u$ and $\vecmy{x_i}$ denote the raw user and item features (Figure \ref{fig:method}). Under this setup, the Proportional Odds Model (Equation (\ref{eq:pom})) is reformulated as: 
{\small\begin{equation}
  \log \frac{P(\rv{k} \le c\mid\vecmy{x}_u,\vecmy{x}_i)}{P(\rv{k} > c\mid\vecmy{x}_u,\vecmy{x}_i)}=a_c-\mathcal{K}(\vecmy{e}_u,\vecmy{e}_i),\notag
\end{equation}}

\smallskip\noindent
where $\mathcal{K}$ is a kernel to measure similarity. 

\subsubsection{Nested Optimization Framework} With Neural OLR for greater flexibility, we now address its remaining limitation: \textbf{shared encoders} still restrict the model’s ability to learn feedback-specific feature dependencies (Figure~\ref{fig:method}). A naive solution—training completely separate encoders for each feedback type—would fragment the embedding space. To address this issue, we propose a \textbf{Nested Optimization Framework} with two key components: \textit{Nested Category-Specific Encoding} and \textit{Nested OLR Optimization}: 

\textbf{(1) Nested Category-Specific Encoding} is designed for category-aware feature extraction with separate encoders. Instead of generating a single user preference embedding directly, we construct a sequence of user-item embedding subspaces using multiple Twin Tower encoders. Each encoder captures a distinct level of user engagement, contributing to a progressively enriched representation.

Formally, consider we have $T$ implicit feedback types, mapped into an ordinal category sequence $[0,1,\dots,T+1]$. Let $(f_u^c,f_i^c)$ represent the neural encoders for each category $c\in [1,T]$, producing the embeddings: $(\vecmy{e}_u^c,\vecmy{e}_i^c)=(f_u^c(\vecmy{x}_u^c),f_i^c(\vecmy{x}_i^c))$ and are $\ell_2$-normalized. To get a \textbf{unified space}, we define the \textbf{Nested Embedding} as:
\begin{equation}
    \vecmy{E}_u^{c}=Concat[\vecmy{e}_u^{1},\dots , \vecmy{e}_u^{c}],\quad \vecmy{E}_i^{c}=Concat[\vecmy{e}_i^{1},\dots , \vecmy{e}_i^{c}] \notag
\end{equation}
Each $\vecmy{E}_u^{c}$ and $\vecmy{E}_i^{c}$ aggregates representations from all preceding levels, ensuring all lower-order information is considered in predicting higher order categories. Note that category $0$ and $T+1$ do not require learned embeddings. Building upon nested embeddings, we generalize the proportional odds model (Equation~(\ref{eq:pom})) as follows:
\begin{equation}
    \log \frac{P(\rv{k} \le c\mid\vecmy{x})}{P(\rv{k} > c\mid\vecmy{x})}=a_c-c\gamma\mathcal{K}(\vecmy{E}_u^c,\vecmy{E}_i^c),
\end{equation}
where $\mathcal{K}(\vecmy{a},\vecmy{b})=\frac{\vecmy{a}^T\vecmy{b}}{\|\vecmy{a}\|\cdot\|\vecmy{b}\|}$ (\emph{Cosine}) and $\gamma$ is a reshaping factor controlling the output distribution of the kernel function. Following the cumulative probability formulation in Equation~(\ref{eq:cdf}), we derive:
\begin{equation}
    P(\rv{k} \le c \mid \vecmy{x}_u, \vecmy{x}_i) = \sigma(a_c - c\gamma\mathcal{K}(\vecmy{E}_u^c, \vecmy{E}_i^c)),
\label{eq:cdf2}
\end{equation}
and thus the probability of predicting category $k=c$ becomes:
{\small
\begin{equation}
    P(\rv{k} = c | \vecmy{x}_u,\vecmy{x}_i) = \sigma(a_c-c\gamma\mathcal{K}(\vecmy{E}_u^c,\vecmy{E}_i^c)) - \sigma(a_{c-1}-(c-1)\gamma\mathcal{K}(\vecmy{E}_u^{c-1},\vecmy{E}_i^{c-1})) 
\label{eq:pdf2}
\end{equation}
}
Here, $a_c$ and $a_{c-1}$ act as constants in the OLR paradigm. By leveraging the monotonicity of the sigmoid ($\sigma$), maximizing $P(\rv{k}=c\mid\vecmy{x}_u,\vecmy{x}_i)$ effectively pushes $\mathcal{K}(\vecmy{E}_u^{c-1},\vecmy{E}_i^{c-1})$ and $\mathcal{K}(\vecmy{E}_u^{c},\vecmy{E}_i^{c})$ apart. If $\vecmy{e}_u^c$ and $\vecmy{e}_u^c$ are $\ell_2$-normalized and $\mathcal{K}$ is cosine, then when the user’s preference level truly reaches category $c$, the model promotes \textbf{similarity} in the preceding sub-embeddings $\vecmy{E}_u^{c-1}$ and $\vecmy{E}_i^{c-1}$ (lower-level categories) and enforces \textbf{dissimilarity} in $\vecmy{e}_u^c$ and $\vecmy{e}_i^c$. Consequently, $\vecmy{e}_u^c$ and $\vecmy{e}_i^c$ contribute positively only if the user reaches a higher preference level than $c$, effectively acting as a "goalkeeper". In other words, each higher-level embedding is “activated” only after the user preference has surpassed the corresponding threshold.

Further, note that $P(\rv{k}>T\mid\vecmy{x}_u,\vecmy{x}_i)$ can be written as $1-P(\rv{k}\le T\mid\vecmy{x}_u,\vecmy{x}_i)$ $=1-\sigma(a_T-\gamma\sum_{c=1}^T\mathcal{K}(\vecmy{e}_u^c,\vecmy{e}_i^c))$, implying an aggregation of all preference-level information. Thus $P(\rv{k}>T\mid\vecmy{x}_u,\vecmy{x}_i)$ is also interpreted as the \textbf{unified preference score}, and $\vecmy{E}_u^T,\vecmy{E}_i^T$ are the \textbf{unified preference embeddings}, where higher similarity between them indicates a higher preference level. 

\textbf{(2) Nested OLR Optimization.} Standard OLR assigns each sample to exactly one category. However, maximizing $P(\rv{k}=c)$ alone does not ensure a progressive probability distribution for the preceding categories. For instance, consider three categories ($k\le 3$) and following scenarios. Scenario A: $P(\rv{k}=1)=0.4, P(\rv{k}=2)=0,P(\rv{k}=3)=0.6$ and Scenario B: $P(\rv{k}=1)=0.1,P(\rv{k}=2)=0.3,P(\rv{k}=3)=0.6$. They all produces $P(\rv{k}=3)=0.6$ but only Scenario B reflects a user's ideal preference distribution who just surpasses preference level 2 but not exceeds level 3.

To stabilize learning and ensure each category’s probability aligns with ideal user’s engagement progression, we propose a Nested OLR Optimization framework. Specifically, we define $T$ subtasks, each focusing on a partial view of the category space. For subtask $t\le T$, we consider only categories 
$\{0,\dots,t+1\}$. Any sample with a assigned category $k< t+1$ retains its label, while any sample with a label $k\ge t+1$ is re-labeled to $t+1$. This merges higher categories into one, ensuring subtask $t$ focuses probability distribution among the preceding levels ($k\le t +1$), namely \textit{category remapping}. Let $k_s^t$ be the re-mapped category of sample $s$ for subtask $t$. Using the probabilities from Equation~(\ref{eq:pdf2}), the overall loss is:

\noindent $
L=\sum_{t=1}^{T} L_t, \quad L_t=-\sum_{s=1}^S\sum_{c=1}^{t+1}I(k_s^{t}=c)\log P(\rv{k}_s^{t}=c\mid\vecmy{x}_u^s,\vecmy{x}_i^s), 
$

\noindent By optimizing all subtasks jointly, the model is encouraged to distribute probabilities appropriately across all categories. We refer to this entire framework as \textbf{Generalized Neural OLR (GNOLR)}.

\subsection{Compare with Feedback-Wise Modeling}
\label{sec:theory}
Predominant approaches \cite{ESMM,ESCM2,DCMT,TAFE,NISE} commonly model multiple implicit feedback signals by optimizing a set of binary classification tasks using \textbf{Cross-Entropy (CE)} loss. In this section, we theoretically demonstrate why GNOLR offers advantages over these approaches by comparing GNOLR with standard CE.

\begin{figure}[t]
    \centering
    \includegraphics[width=1\linewidth]{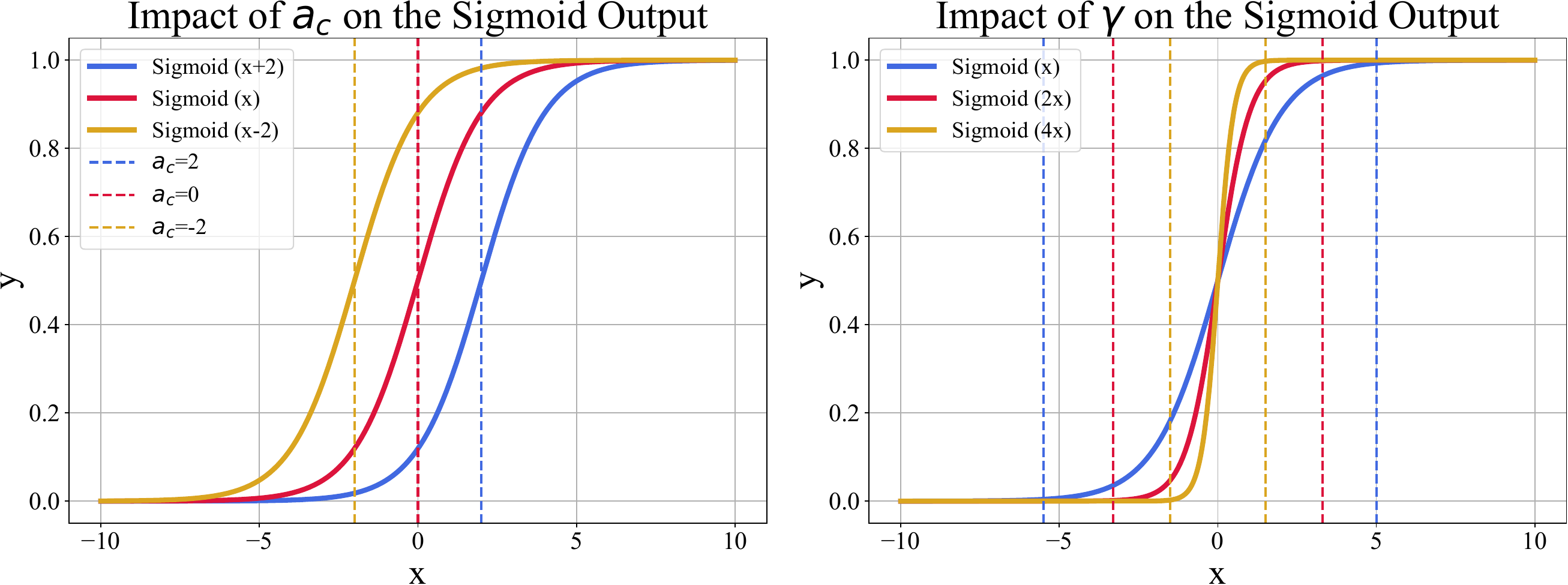}
    \caption{The impact of $a_c$ and $\gamma$ on Sigmoid predictions. $a_c$ shifts the Sigmoid curve horizontally. $\gamma$ modifies the steepness of the Sigmoid curve, controlling which region of the input space receives greater focus during learning.}
    \Description[Effect of ac and gamma on Sigmoid curve]{This figure illustrates how the parameters \( a_c \) and \( \gamma \) affect the Sigmoid function. Increasing or decreasing \( a_c \) shifts the curve left or right along the x-axis, while changing \( \gamma \) alters the steepness of the curve. A larger \( \gamma \) results in a steeper transition, focusing learning on a narrower input range.}
    \label{fig:sigmoid}
\end{figure}

\subsubsection{Single Feedback Case} Here we show GNOLR’s mathematical equivalence to CE loss while $\{a_c\}$ and $\gamma$ enhance adaptability. Consider we only have clicks as feedback. Through mapping, we derive three categories $k\in \{0,1,2\}$ where $k=1$ denotes no interaction, and $k=2$ is a click. By definition, $I(k=1)=1-I(k=2)$. Under this setup, the GNOLR loss can be expressed as: $L=-\sum_{s=1}^S\sum_{c=1}^2 I(k_s=c) \log P(\rv{k}_s=c\mid\vecmy{x}_u^s,\vecmy{x}_i^s)$. Expand $P(\rv{k}_s=c)$ and let $y=I(k_s=2)$, the loss for sample $s$ becomes:

\noindent$L=-(1-y)\log \sigma(a_1-\gamma\mathcal{K}(\vecmy{e}_u,\vecmy{e}_i))-y\log (1-\sigma(a_1-\gamma\mathcal{K}(\vecmy{e}_u,\vecmy{e}_i)))$
% \smallskip

\noindent Rewriting with $\sigma(x)=1-\sigma(-x)$, we have:

\noindent$L= -(1-y)\log (1-\sigma(\gamma\mathcal{K}(\vecmy{e}_u,\vecmy{e}_i)-a_1))-y\log (\sigma(\gamma\mathcal{K}(\vecmy{e}_u,\vecmy{e}_i)-a_1))$. 

\noindent This is equivalent to the CE loss. However, the key distinction is that GNOLR introduces two key parameters $a_1$ and $\gamma$ which changes the output distribution of the Cosine kernel. From the perspective of CE loss, the threshold $a_1$ ensures that the model predicts high probability only when $\gamma\mathcal{K}(\vecmy{e}_u,\vecmy{e}_i)$ exceeds $a_1$. This mechanism effectively pulls user and item embeddings closer for positive pairs, enhancing nearest neighbor retrieval. Additionally, $\gamma$ controls the scaling of $\mathcal{K}(\vecmy{e}_u,\vecmy{e}_i)$, thereby influencing the range of inputs that receive significant gradient updates. As a result, adjusting $\gamma$ shifts the model’s focus toward different sample distributions, improving adaptability to varying levels of difficulty in the data.

This reformulation reveals that GNOLR enhances model's flexibility and adaptability over standard CE by shifting and reshaping the $\sigma$'s output distribution with $a_c$ and $\gamma$. The impact of $a_c$ and $\gamma$ on the prediction distribution is illustrated in Figure~\ref{fig:sigmoid}.

\subsubsection{Multi-Feedback Case} Here we show that GNOLR avoids contradictory label assignments in multi-feedback CE loss by modeling multi-level feedback in a unified and consistent manner. For simplicity, consider a scenario with two feedback: clicks and purchases (pay). We define a category list $k\in[0,1,2,3]$, where $k=0$ is null, $k=1$ means no feedback, $k=2$ means clicked but unpaid, $k=3$ means paid. Through the eyes of prior multi-task frameworks \cite{ESMM,NISE,TAFE}, we can interpret the category indicators as: $I(k=1)=1-y_{click}$, $I(k=2)=\max(y_{click}-y_{pay},0)$, and $I(k=3)=y_{pay}$.
Here, $y_{click}$ and $y_{pay}$ are binary indicators for click and purchase, respectively. Similarly, we can infer the predicted probabilities for each category as: $p(\rv{k}=1\mid\vecmy{x})=1-p_{ctr}$, $p(\rv{k}=3\mid\vecmy{x})=p_{ctcvr}$, and thus $p(\rv{k}=2\mid\vecmy{x})=p_{ctr}-p_{ctcvr}$, where $p_{ctr}$ is the predicted Click-Through Rate (CTR), and $p_{ctcvr}$ is the predicted Purchase Rate (CTCVR). The loss function for a single sample under GNOLR can then be expands to:
{\small\begin{align}
    &L=-2(1-y_{click})\log (1-p_{ctr}) -y_{click}\log p_{ctr} -y_{pay}\log p_{ctcvr} \notag \\
    & \quad\quad -\max(y_{click}-y_{pay},0)\log (p_{ctr}-p_{ctcvr})
\label{eq:case2}
\end{align}}

\noindent In feedback-wise approaches (using separate CE loss for each feedback)~\cite{ESMM,ESCM2,DCMT,TAFE,NISE}, a CTR head is used to predict $p_{ctr}$, and a CTCVR head predicts $p_{ctcvr}$. Although the first line in Equation (\ref{eq:case2}) appear in feedback-wise CE loss, a crucial difference lies in how clicked and unpurchased samples ($y_{click}=1,y_{pay}=0$) are treated. CE loss uniformly pushes $p_{ctcvr}$ to 0 for all unpurchased samples, while increasing $p_{ctr}$ to 1 for all clicked items, which may lead to inconsistent supervision signals between the CTR and CTCVR heads. By contrast, GNOLR avoids explicitly assigning such conflicting targets to clicked-but-unpurchased items, and instead emphasizes the relative difference between $p_{ctr}$ and $p_{ctcvr}$, potentially leading to more coherent learning dynamics.

\subsection{Manual Ordinal Thresholds Selection}
\label{sec:cate-threshold}
In conventional OLR, the thresholds $\{a_c\}$ follow an inherent order: $a_1<a_2<...<a_T$ and can be learned jointly with the model. However, GNOLR's nested optimization framework may disrupt this ordering. To resolve this, we propose a novel approach to directly compute a near-optimal ordinal set $\{a_c\}$, eliminating the need for hyperparameter tuning. Specifically, we calculate $\{a_c\}$ by: $a_c\approx \log\frac{1-\mathbb{E}[P(\rv{k}>c\mid\vecmy{x})]}{\mathbb{E}(P(\rv{k}>c\mid\vecmy{x}))}$, where $\mathbb{E}[P(\rv{k}>c\mid\vecmy{x})]$ denotes the sample proportion of instances belonging to categories strictly above $c$. This quantity is straightforward to compute from the data.

Specifically, given that cosine similarity is symmetric and bounded in [-1,1], we encourage $\mathbb{E}_{\vecmy{x}}[\mathcal{K}(\vecmy{E}_u,\vecmy{E}_i)]\approx 0$ for balanced similarity distribution. For category $c$, the OLR-based formulation implies $\log \frac{1-P(\rv{k}>c\mid\vecmy{x})}{P(\rv{k}>c\mid\vecmy{x})} = a_c-c\gamma\mathcal{K}(\vecmy{E}_{u}^c,\vecmy{E}_i^c)$. Taking expectations over $\vecmy{x}$ and applying Jensen’s inequality for approximation gives:

\smallskip\noindent
$\quad \mathbb{E}[a_c-c\gamma\mathcal{K}(\vecmy{e}_{u},\vecmy{e}_i)]=\mathbb{E}[\log \frac{1-P(\rv{k}>c|\vecmy{x})}{P(\rv{k}>c|\vecmy{x})}]\geq \log\mathbb{E}[\frac{1-P(\rv{k}>c|\vecmy{x})}{P(\rv{k}>c|\vecmy{x})}]$

\smallskip\noindent
$\quad\quad\quad\quad a_c\approx \log\mathbb{E}[\frac{1-P(\rv{k}>c|\vecmy{x})}{P(\rv{k}>c|\vecmy{x})}]\approx\log \frac{1-\mathbb{E}[P(\rv{k}>c|\vecmy{x})]}{\mathbb{E}(P(\rv{k}>c|\vecmy{x}))}$,
\smallskip

Because our label mapping orders implicit feedback signals in ascending order of sparsity, the thresholds $\{a_c\}$ estimated via this approach naturally form an increasing sequence. In practice, \textbf{manually setting these thresholds often stabilizes training}, and we observe that thresholds obtained via learning are typically close to the values derived from the above empirical approximation.
\section{EXPERIMENTS}
In the context of embedding-based collaborative filtering, an effective algorithm should achieve two objectives: \textbf{(1) align spatial proximity with user interest}, ensuring that similar user-item pairs are closer in the embedding space, and \textbf{(2) provide strong personalized ranking capabilities}, displaying items with higher engagement propensity at higher-ranked positions in the list. In this section, we conduct extensive experiments on nine public real-world datasets to address the following research questions: 
\begin{itemize}[leftmargin=0em,itemindent=1em,labelsep=0.5em]
    \item \textbf{RQ1} How does GNOLR perform on single-task and multi-task ranking compared to state-of-the-art (SOTA) methods?
    \item \textbf{RQ2} How does GNOLR perform on the embedding-based collaborative filtering (retrieval) task relative to the baselines?
    \item \textbf{RQ3} How does GNOLR deal with prevalent listwise modeling?
    \item \textbf{RQ4} How sensitive is GNOLR to the hyper-parameter settings?
    \item \textbf{RQ5} How much does the Nested Optimization Framework contribute to GNOLR's performance?
\end{itemize}

\begin{table}[t]
\small
\centering
\caption{Statistics of all datasets. The number of users and items in the AE series datasets is not disclosed. \#Pos$_1$ and \#Pos$_2$ denotes the counts for click and pay in e-commercial datasets, likes and follows for video datasets.}
\label{tab:datasets}
\begin{tabular}{cccccc}
\hline
\hline
Dataset & \#User & \#Item & Total  & \#Pos$_1$ & \#Pos$_2$ \\ \hline
Ali-CCP & 0.24M  & 0.47M  & 69.1M  & 2.62M     & 13.1K     \\
AE-ES   & /      & /      & 31.67M & 0.84M     & 19.1K     \\
AE-FR   & /      & /      & 27.04M & 0.54M     & 14.4K     \\
AE-NL   & /      & /      & 7.31M  & 0.17M     & 6K        \\
AE-US   & /      & /      & 27.39M & 0.45M     & 10.8K     \\
KR-Pure & 24.9K  & 6.8K   & 0.92M  & 19.78K    & 1.13K     \\
KR-1K   & 1K     & 2.4M   & 6.75M  & 0.13M     & 7.98K     \\
ML-1M   & 6K     & 3.7K   & 1M     & 0.23M     & /         \\
ML-20M  & 138K   & 27K    & 20M    & 4.43M     & /         \\
RetailR & 1.4M   & 0.24M  & 2.76M  & 91.8K     & 22.5K     \\ \hline
\hline
\end{tabular}
\end{table}

\subsection{Experiment Setup}
\subsubsection{Datasets}
\label{sec:dataset}
To comprehensively evaluate GNOLR, we use nine widely-adopted, large-scale, and real-world datasets spanning diverse recommendation scenarios and task settings: \textbf{(1) Ali-CCP} \cite{ESMM}: An e-commerce dataset from Taobao’s recommender system, containing two implicit feedback types (click and pay). To align with the collaborative filtering setup, we discard user-item cross features and use only user-side and item-side features. \textbf{(2) AE} \cite{AE}: An e-commerce dataset from AliExpress’s search logs, with four sub-datasets from different country markets (\textbf{AE-ES, AE-FR, AE-NL, AE-US}). Each sub-dataset includes click and pay feedback. \textbf{(3) KuaiRand} \cite{KuaiRand}: A video recommendation dataset from the Kuaishou app. We use two versions \textbf{(KR-Pure, KR-1K)} of it, differing by their user and item sampling strategies. Two types of implicit feedback, likes and follows, are used in modeling. \textbf{(4) RetailRocket}: A dataset from a real-world e-commerce website, containing three types of user behaviors: click, add-to-cart (ATC), and Pay. We use the latter two. \textbf{(5) MovieLens} \cite{MovieLens}: A widely used movie recommendation dataset, with two versions (\textbf{ML-1M, ML-20M}). We set a rating threshold ($> 4$) to create binary labels to simulate implicit feedback. To simulate real-world scenarios, we sort all samples chronologically and split the data into 70\% (history) for training and 30\% (future) for testing. The statistics are listed in Table~\ref{tab:datasets}.

\begin{table}[t]
\caption{Single-Task Ranking Results (AUC). Methods predict the probability of clicks for Ali-CCP and AE, predict likes for KuaiRand, ATC for RetailRocket, and predict positive for the MovieLens datasets. The best results (statistically significant) are highlighted in bold, while the second-best results are underlined.}
\centering
\small
\begin{tabular}{cccccc}
\hline
\hline
Method  & BCE    & BCE$^*$       & JRC    & JRC$^*$ & GNOLR \\
\hline
AliCCP  & 0.5005 & 0.5896        & 0.5009 & {\ul 0.5924}  & \textbf{0.6232} \\
AE-ES   & 0.5078 & 0.7118        & 0.6323 & {\ul 0.7357}  & \textbf{0.7366} \\
AE-FR   & 0.5008 & 0.6887        & 0.6428 & {\ul 0.7307}  & \textbf{0.7335} \\
AE-US   & 0.5078 & 0.6165        & 0.6256 & {\ul 0.6831}  & \textbf{0.7062} \\
AE-NL   & 0.5155 & 0.6715        & 0.6419 & {\ul 0.6853}  & \textbf{0.7298} \\
KR-Pure & 0.6056 & 0.7665        & 0.7835 & {\ul 0.8280}  & \textbf{0.8506} \\
KR-1K   & 0.7485 & 0.8510        & 0.8018 & {\ul 0.8684}  & \textbf{0.9024} \\
ML-1M   & 0.7542 & 0.7896        & 0.7937 & {\ul 0.7939}  & \textbf{0.8139} \\
ML-20M  & 0.7535 & 0.7790        & 0.7801 & {\ul 0.7854}  & \textbf{0.8094} \\
RetailR & 0.5034 & {\ul 0.7308}  & 0.5009 & 0.7165        & \textbf{0.7537} \\
\hline
\hline
\end{tabular}
\label{tab:single-task-pointwise}
\end{table}

\begin{table}[t]
\centering
\caption{The GAUC performance comparison of various ranking methods across multiple datasets. The best results (statistically significant) are highlighted in bold, while the second-best are underlined. $\lambda$Rank is LambdaRank.}
\label{tab:single-task-listrank}
\resizebox{\columnwidth}{!}{
\begin{tabular}{cccccccccc}
\hline
\hline
Dataset       & RankNet      & $\lambda$Rank   & ListNet      & S2SRank  & SetRank         & JRC          & GNOLR$_L$      \\ \hline
AliCCP        & 0.5518       & {\ul 0.5534} & 0.5447       & 0.5499       & 0.5523          & 0.5009       & \textbf{0.5602} \\
AE-ES         & {\ul 0.5432} & 0.5426       & 0.5431       & 0.5421       & 0.5403          & 0.5236       & \textbf{0.5433} \\
AE-FR         & {\ul 0.5351} & 0.5343       & 0.5338       & 0.5347       & 0.5331          & 0.5159       & \textbf{0.5364} \\
AE-US         & {\ul 0.5294} & 0.5291       & 0.5289       & 0.5287       & 0.5273          & 0.5159       & \textbf{0.5300} \\
AE-NL         & {\ul 0.5278} & 0.5270       & 0.5272       & 0.5275       & 0.5274          & 0.5201       & \textbf{0.5346} \\
KR-Pure       & {\ul 0.5090} & 0.5073       & 0.5075       & 0.5079       & \textbf{0.5106} & 0.5082       & 0.5084          \\
KR-1K         & 0.5062       & 0.4953       & 0.5084       & 0.5136       & {\ul 0.5242}    & 0.5036       & \textbf{0.5380} \\
ML-1M         & {\ul 0.7121} & 0.6891       & 0.6985       & 0.7057       & 0.6974          & 0.7040       & \textbf{0.7244} \\
ML-20M        & 0.6941       & 0.6869       & {\ul 0.6969} & 0.6874       & 0.6730          & 0.6817       & \textbf{0.7006} \\
\hline
\hline
\end{tabular}}
\end{table}

\subsubsection{Evaluation Metrics}
We use the following metrics for different problems. For RQ1, RQ3, and RQ4, we use the widely adopted \textbf{AUC} (Area Under the ROC Curve) \cite{ESMM,ESCM2,DCMT,NISE,TAFE}. For RQ3, we add the \textbf{GAUC} (AUC averaged over user sessions) \cite{he2016ups,DIN} to measure listwise performance. For RQ2, we use \textbf{Recall@K} \cite{EBR,zhang2016collaborative,wang2019knowledge}, where each user’s candidate items are ranked by Euclidean distance between user and item embeddings, and the proportion of positive items (of corresponding targets) in the top $K$ is calculated.

\subsubsection{Baselines} We use state-of-the-art methods as baselines across different task settings: \textbf{For RQ1}, we consider Binary Cross Entropy (BCE) and JRC \cite{JRC} as single-task baselines. For multi-task settings, we include NSB (Naive Shared Bottom), ESMM \cite{ESMM}, ESCM\textsuperscript{2}-IPS \cite{ESCM2}, ESCM\textsuperscript{2}-DR \cite{ESCM2}, DCMT \cite{DCMT}, NISE \cite{NISE}, and TAFE \cite{TAFE}. \textbf{For RQ2}, all pointwise, pairwise, listwise, or setwise ranking methods are suitable for \textbf{single-task} personalized retrieval, as they all focus on distinguishing positive samples from negatives for each user. We evaluate BCE, RankNet \cite{RankNet}, LambdaRank \cite{LambdaRank}, ListNet \cite{ListNet}, SetRank \cite{SetRank}, S2SRank \cite{Set2setRank}, and JRC \cite{JRC}. However, for \textbf{multi-task} personalized retrieval, none of the existing SOTA multi-task methods can generate retrieval-oriented embeddings for all tasks. For instance, ESMM models ctcvr as the product of ctr and cvr, thus the embeddings from the cvr encoder cannot be used for ctcvr-oriented retrieval. Therefore, we only compare GNOLR with NSB, as it directly associates its embeddings with corresponding labels. \textbf{For RQ3}, we include: RankNet \cite{RankNet}, LambdaRank \cite{LambdaRank}, ListNet \cite{ListNet}, SetRank \cite{SetRank}, S2SRank \cite{Set2setRank}, and JRC \cite{JRC}.

Methods marked with "*" indicate that they were fine-tuned with a \textbf{sample reweighting} technique \cite{guo2022learning,liu2022improving}. Specifically, we only increase the weight of positive samples for each feedback.

\begin{table*}[t]
\centering
\caption{Multi-Task Ranking Results (AUC). The best results are highlighted in bold, while the second-best are underlined.}
\label{tab:multi-task}
\begin{tabular}{ccccccccccc}
\hline
\hline
Task                    & Dataset       & NSB$^*$ & ESMM$^*$     & ESCM\textsuperscript{2}-IPS$^*$ & ESCM\textsuperscript{2}-DR$^*$ & DCMT$^*$     & NISE$^*$     & TAFE$^*$     & Neural OLR      & GNOLR           \\ \hline
\multirow{5}{*}{CTR}    & AliCCP        & 0.5877  & 0.5955       & 0.5947                          & 0.5962                         & 0.5952       & 0.6020       & 0.5981       & \textbf{0.6162} & {\ul 0.6153}    \\
                        & AE-ES         & 0.7024  & 0.7084       & {\ul 0.7337}                    & 0.7271                         & 0.6963       & 0.7079       & 0.7089       & 0.7257          & \textbf{0.7372} \\
                        & AE-FR         & 0.7103  & 0.6798       & {\ul 0.7322}                    & 0.7156                         & 0.6714       & 0.6932       & 0.6959       & 0.7194          & \textbf{0.7370} \\
                        & AE-US         & 0.6415  & 0.6593       & 0.6837                          & \textbf{0.7014}                & 0.6513       & 0.6735       & 0.6751       & 0.6792          & {\ul 0.6971}    \\
                        & AE-NL         & 0.6505  & 0.6903       & 0.6656                          & 0.6706                         & 0.6621       & 0.6882       & {\ul 0.6975} & 0.6563          & \textbf{0.7277} \\ \hline
\multirow{5}{*}{CTCVR}  & AliCCP        & 0.5070  & 0.5453       & 0.5439                          & {\ul 0.5527}                   & 0.5412       & 0.5351       & 0.5336       & 0.5362          & \textbf{0.5997} \\
                        & AE-ES         & 0.7150  & 0.8433       & 0.8439                          & {\ul 0.8563}                   & 0.8483       & 0.8143       & 0.8139       & 0.7858          & \textbf{0.8827} \\
                        & AE-FR         & 0.7262  & 0.8219       & 0.8274                          & 0.8565                         & {\ul 0.8610} & 0.7797       & 0.7874       & 0.7618          & \textbf{0.8793} \\
                        & AE-US         & 0.6850  & 0.8237       & 0.7820                          & {\ul 0.8403}                   & 0.8221       & 0.7719       & 0.7820       & 0.7354          & \textbf{0.8663} \\
                        & AE-NL         & 0.7143  & {\ul 0.8129} & 0.7808                          & 0.7986                         & 0.8051       & 0.7601       & 0.7801       & 0.7379          & \textbf{0.8343} \\ \hline
\multirow{2}{*}{Like}   & KR-Pure       & 0.8077  & 0.8157       & 0.8077                          & 0.8124                         & 0.8250       & 0.8271       & 0.8286       & {\ul 0.8440}    & \textbf{0.8456} \\
                        & KR-1K         & 0.8865  & 0.8817       & 0.8892                          & 0.8858                         & 0.8794       & 0.8816       & 0.8890       & {\ul 0.9079}    & \textbf{0.9087} \\ \hline
\multirow{2}{*}{Follow} & KR-Pure       & 0.7025  & 0.6994       & 0.6909                          & 0.7025                         & 0.7056       & 0.6994       & 0.6963       & \textbf{0.7212} & {\ul 0.7161}    \\
                        & KR-1K         & 0.7318  & 0.7399       & {\ul 0.8113}                    & 0.7888                         & 0.7999       & 0.6868       & 0.7327       & 0.7100          & \textbf{0.8165} \\ \hline
\multirow{1}{*}{ATC}    & RetailR       & 0.6845  & 0.6981       & 0.6912                          & 0.6926                         & 0.6942       & 0.6945       & 0.7013       & {\ul 0.7521}    & \textbf{0.7764}    \\ 
\multirow{1}{*}{Pay}    & RetailR       & 0.7247  & 0.8148       & 0.8209                          & 0.8139                         & 0.8150       & {\ul 0.8238} & 0.7868       & 0.8189          & \textbf{0.8242}    \\ \hline
\hline
\end{tabular}
\end{table*}

\subsubsection{Implementations}
To ensure a fair comparison, all methods use the same backbone network architecture: \textbf{Twin Tower encoders with standard MLPs} \cite{DSSM}. Additionally, all embeddings are \textbf{$\ell_2$-normalized} to facilitate compatibility with nearest neighbor search in the Euclidean space \cite{GoogleRetrieval,EBR}. Hyper-parameters for each method are tuned independently (grid-search) for each dataset to achieve optimal performance (see Appendix for details). The reported results represent the average of ten independent runs. \textbf{Our code} is available at \url{https://github.com/FuCongResearchSquad/GNOLR}.

\subsection{Results and Analysis}
\subsubsection{\textbf{RQ1: Ranking Performance}}
Table~\ref{tab:single-task-pointwise} and Table~\ref{tab:multi-task} present the ranking performance of various methods under single-task and multi-task settings. We have the following observations: 

\noindent\textbf{(1) Overall Gains.} GNOLR demonstrates superior performance consistently across both single-task and multi-task settings. Its ability to model complex user preferences and leverage additional implicit signals effectively is evident from its significant improvements over the baselines. Notably, the progression of user engagement is captured effectively in different recommendation contexts (e.g., e-commerce, movie, and short video recommendation).

\noindent\textbf{(2) Robustness to Extremely Imbalanced Distribution.} GNOLR maintains strong performance on highly imbalanced e-commerce datasets (e.g., Ali-CCP, AE), where baselines often fail (AUC $\approx 0.5$). This is due to cross-entropy loss struggles with such imbalanced distributions~\cite{focalloss}. While \textbf{sample reweighting} improves baselines' performance, GNOLR outperforms them \textbf{without} requiring such adjustments. This advantage stems from GNOLR’s structured label hierarchy, which models feedback progression instead of dealing with them independently. Additionally, GNOLR sets ${a_c}$ directly based on label distribution, avoiding tuning, while $\gamma$ enables hard sample mining. These hyperparameters adaptively adjust GNOLR's prediction distributions and further enhance robustness.

\noindent\textbf{(3) Prior Limitation in Task Balancing.} Even after applying sample re-weighting, the baselines remain inferior to GNOLR. This can be attributed to the \textbf{potential conflicts} induced by the feedback-wise formulation, which increase the difficulty of finding a robust Pareto frontier of task-specific weights for prior frameworks~\cite{MPE}. Consequently, these methods suffer from \textbf{gradient conflicts} arising from improper label modeling, as discussed in Section~\ref{sec:theory}.

\noindent\textbf{(4) Impact of Minority Category.} GNOLR shows minor performance variations when transitioning from single-task to multi-task prediction. For instance, the AUC for click prediction on AliCCP drops from 0.6232 to 0.6153, while it improves from 0.7335 to 0.7370 on AE-FR. Generally, we expect GNOLR to improve as more implicit feedback signals are incorporated, as they provide richer information about user preferences. However, the minor performance drop in some cases, such as AliCCP, may be attributed to the sparsity of certain interactions, since GNOLR shows uplift on dense-interacted datasets like AE-ES and AE-FR. Sparse positive feedback can lead to \textbf{insufficient distribution fitting} near the ordinal category boundaries ($\{a_c\}$). Addressing this limitation is left for future work.

\subsubsection{\textbf{RQ2: Retrieval Performance}}
The results of single- and multi-task retrieval are presented in Tables~\ref{tab:recall-ml-1m} and~\ref{tab:recall-kuairand}. GNOLR demonstrates superior performance in both single-task and multi-task retrieval tasks. Key observations are as follows.

\noindent\textbf{(1) Adaptability to Imbalanced Data.} GNOLR’s advantage in single-task retrieval arises from its adaptability to imbalanced distributions. The hyperparameter $\gamma$ reshapes the sigmoid curve to emphasize challenging samples, while $a_c$ shifts the curve, refining the boundary between positive and negative items. These adjustments yield a more coherent embedding space topology with the data distribution and thus stronger retrieval performance.

\noindent\textbf{(2) Unified Embedding for Multi-Target Retrieval.} In multi-task retrieval, GNOLR significantly outperforms NSB* (NSB with reweighting), especially on sparse targets. Unlike NSB*’s task-specific embeddings, GNOLR employs a unified embedding space, enabling retrieval across multiple feedback signals with \textbf{the same set of embeddings}. Conventional methods fail to capture label correlations, particularly the progressiveness of user interest. This results in disjoint embedding spaces and lower recall when applying one task’s embedding to another, e.g., evaluating embedding NSB$_{\text{follow}}$ on the "like" target will lead to drop in Recall. GNOLR’s unified framework avoids this pitfall by leveraging the \textbf{category-specific nested optimization} framework (Figure \ref{fig:method}) among tasks.

\noindent\textbf{(3) Spatial Structure Visualization.} 
The superiority of GNOLR in retrieval is illustrated in the visualization (Figure~\ref{fig:case-study}). For single-task settings, embeddings trained with standard CE loss exhibit a poor discriminative topology: the angle between the user vector and positive item vectors often exceeds $90^{\circ}$, making nearest neighbor retrieval less effective. In contrast, GNOLR adjusts the angle distribution through $a_c$, ensuring positive items form smaller angles with the user vector, thereby improving retrieval performance.

For multi-task settings, especially with sparse positive labels, CE loss often pushes all items away from the user vector, resulting in a "vacant" region within the half ball centered with the user vector (Figure~\ref{fig:case-study}-2(c)). This compressed embedding space reduces the model’s robustness to unseen data. GNOLR addresses this by shifting decision boundaries among categories towards user vectors according to their preference levels (achieved by proper configuration of $\{a_c\}$), ensuring robustness and improving retrieval performance, particularly for sparse targets.

\begin{table}[t]
\centering
\caption{Single-Task Retrieval on ML-1M. The best result are presented in bold font, while the second-best are underlined.}
\label{tab:recall-ml-1m}
\resizebox{0.96\columnwidth}{!}{
\begin{tabular}{ccccc}
\hline
\hline
Method      & Recall@5        & Recall@10       & Recall@15       & Recall@20       \\ \hline
BCE         & 0.3890          & 0.5842          & 0.6939          & 0.7639          \\
RankNet     & {\ul 0.4011}    & {\ul 0.6007}    & {\ul 0.7110}    & {\ul 0.7807}    \\
LambdaRank  & 0.3924          & 0.5884          & 0.6981          & 0.7672          \\
ListNet     & 0.3903          & 0.5927          & 0.7015          & 0.7704          \\
SetRank     & 0.3931          & 0.5922          & 0.7032          & 0.7725          \\
S2SRank     & 0.3967          & 0.5966          & 0.7077          & 0.7768          \\
JRC         & 0.3684          & 0.5739          & 0.6860          & 0.7579          \\ \hline
GNOLR       & \textbf{0.4086} & \textbf{0.6090} & \textbf{0.7182} & \textbf{0.7865} \\ \hline
\hline
\end{tabular}}
\end{table}

\begin{table}[t]
\centering
\caption{Multi-Target Retrieval on KR-1K. Best in bold font. NSB*$_{\text{like}}$ denotes embeddings from the "like" encoder.}
\label{tab:recall-kuairand}
\resizebox{\columnwidth}{!}{
\begin{tabular}{cccccc}
\hline
\hline
Target                  & Model           & Recall@50       & Recall@100      & Recall@200      & Recall@500      \\ \hline
\multirow{3}{*}{Like}   & NSB*$_{like}$   & 0.0583          & 0.1024          & 0.1809          & 0.3604          \\
                        & NSB*$_{follow}$ & 0.0453          & 0.0911          & 0.1672          & 0.3312          \\
                        & GNOLR           & \textbf{0.0586} & \textbf{0.1085} & \textbf{0.1957} & \textbf{0.3841} \\
                        \hline
\multirow{3}{*}{Follow} & NSB*$_{like}$   & 0.0301          & 0.0614          & 0.1239          & 0.2512          \\
                        & NSB*$_{follow}$ & 0.0331          & 0.0768          & 0.1429          & 0.2710          \\
                        & GNOLR           & \textbf{0.0419} & \textbf{0.0832} & \textbf{0.1529} & \textbf{0.2966} \\ \hline
\hline
\end{tabular}}
\end{table}

\begin{table}[t]
\centering
\caption{Ablation Study on two representative datasets.}
\label{tab:ablation}
\resizebox{\columnwidth}{!}{
\begin{tabular}{cccccc}
\hline
\hline
Dataset                  & Task   & Neural OLR      & GNOLR-V0 & GNOLR-V1 & GNOLR           \\ \hline
\multirow{2}{*}{Ali-CCP} & CTR    & \textbf{0.6114} & 0.6041   & 0.5965   & 0.6085          \\
                         & CTCVR  & 0.5221          & 0.5883   & 0.5854   & \textbf{0.6163} \\ \hline
\multirow{2}{*}{KR-1K}   & Like   & 0.9061          & 0.9023   & 0.9047   & \textbf{0.9079} \\
                         & Follow & 0.7106          & 0.6815   & 0.8076   & \textbf{0.8163} \\ \hline
\hline
\end{tabular}}
\end{table}

\subsubsection{\textbf{RQ3: Co-training with Listwise Loss}} Standard GNOLR primarily falls under pointwise ranking paradigm, but its personalization ability can be further enhanced by combining it with a listwise loss, such as ListNet~\cite{ListNet}. Specifically, the combined loss is defined as $L=L_{GNOLR}+L_{ListNet}$, referred to as GNOLR$_L$. The results are shown in Table~\ref{tab:single-task-listrank}, and the following observations are made: \textbf{(1)} GNOLR$_L$ outperforms the baselines on most datasets, demonstrating its ability to effectively integrate listwise learning into its framework. \textbf{(2)} GNOLR’s strength in pointwise ranking enhances the effectiveness of listwise learning. For example, when combined with ListNet, GNOLR significantly improves both AUC and GAUC compared to ListNet. \textbf{(3)} GNOLR$_L$ achieves substantially better performance than JRC \cite{JRC}, a method specifically designed to balance calibration (AUC) and personalization (GAUC), indicating GNOLR$_L$’s versatility and superior ranking ability in diverse scenarios (See AUC results in Appendix). Note that GAUC is not reported for RetailRocket in the main results, as most users interact with only one item, making personalization effects unobservable; AUC is still provided in the Appendix for reference.

\subsubsection{\textbf{RQ4: Parameter Sensitivity}} 
\begin{figure}[t]
    \centering
    \begin{subfigure}[t]{0.47\columnwidth}
        \centering
        \includegraphics[width=\linewidth]{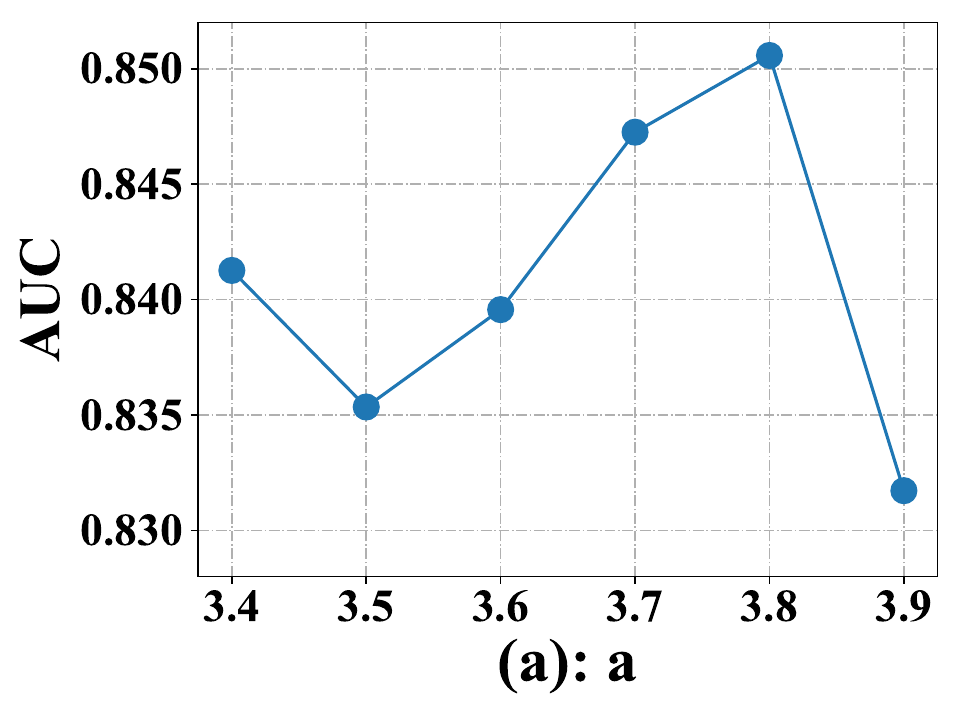}
    \end{subfigure}
    \hfill
    \begin{subfigure}[t]{0.47\columnwidth}
        \centering
        \includegraphics[width=\linewidth]{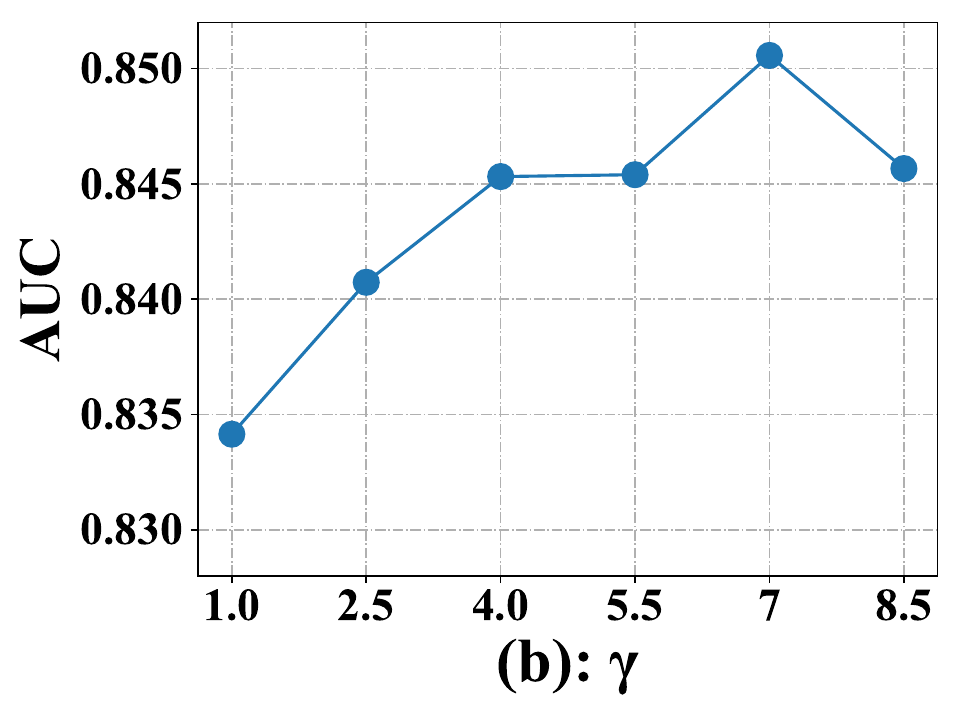}
    \end{subfigure}
    \vspace{-4mm}
    \caption{Parameter sensitivity w.r.t. $a$ and $\gamma$ on KR-Pure.}
    \Description[Impact of hyper-parameter]{This figure illustrates how the parameters \( a_c \) and \( \gamma \) affect GNOLR.}
    \label{fig:hyper}
\end{figure}

We perform parameter sensitivity experiments to analyze the impact of key architectural and optimization parameters on GNOLR's performance. Overall, GNOLR is robust to most hyperparameters, with the exception of $a$ (category thresholds) and $\gamma$ (reshaping factor), which have more impact on performance. The results are presented in Figure~\ref{fig:hyper}. Notably, the optimal value of $a$ aligns closely with the calculation detailed in Section~\ref{sec:cate-threshold}, confirming the theoretical basis for its selection.

\begin{figure}[t]
    \centering
    \includegraphics[width=0.86\linewidth]{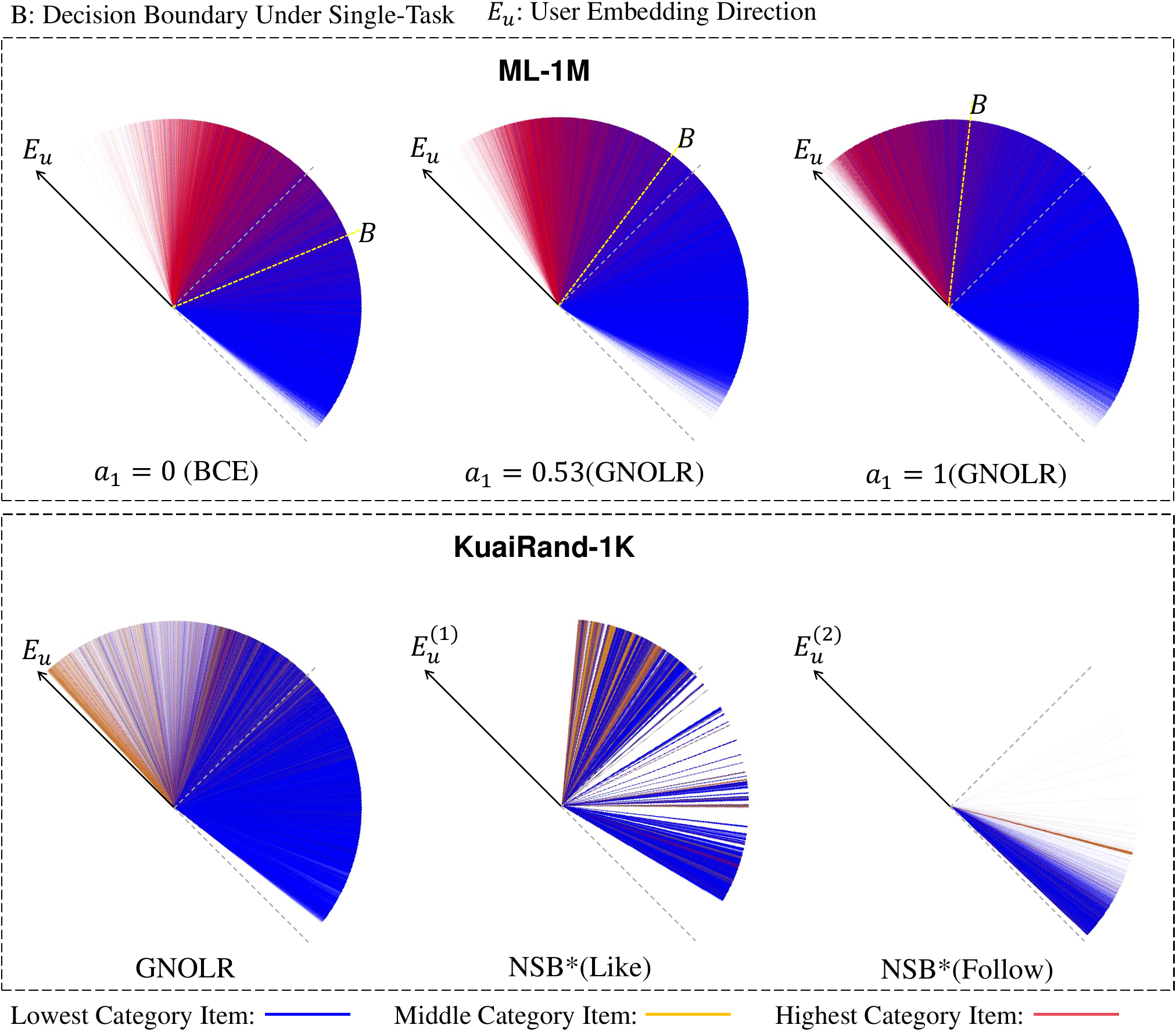}
    \caption{Visualization of the angular distribution between user and item embeddings under single- and multi-task settings. We fix the user directions and plot item directions. NSB* uses sample re-weighting for better performance.} 
    \Description[Visualization of the angular distribution]{Visualization of the angular distribution between user and item embeddings under single- and multi-task settings.}
    \label{fig:case-study}
\end{figure}

\subsubsection{\textbf{RQ5: Ablation Study}} The performance of Naive Neural OLR serves as an ablation baseline to evaluate the impact of GNOLR’s task-specific encoder architecture (Figure \ref{fig:method}). While Naive Neural OLR can occasionally achieve comparable or superior performance on the denser targets, it often suffers from significant degradation on the sparser ones. In contrast, GNOLR consistently delivers balanced performance across all targets. This is attributed to its task-specific encoder design, which selectively models the dependencies between categories and covariates. For Naive Neural OLR, the coefficients are dominated by the correlations between dense label and covariates. Notably, we use a "wider" MLP for Naive Neural OLR to ensure fairness in the total number of parameters.

To further validate the design of GNOLR, we implemented two variants of our GNOLR, i.e., GNOLR-V0 (replaces the shared encoder in Neural OLR with task-specific encoder) and GNOLR-V1 (only incorporating the Nested Category-specific Encoding). As shown in Table~\ref{tab:ablation}, Nested Category-specific Encoding significantly improves performance on sparse targets, and its combination with Nested OLR Optimization achieves the highest gain, demonstrating the effectiveness of the Nested Optimization Framework.

\textbf{Additional results and analyses} are provided in the Appendix.
\section{RELATED WORK}
\textbf{Collaborative filtering (CF)} \cite{CFsurvey,ExplainCF,NCF} leverages collective user feedback to recommend relevant items, traditionally via neighborhood-based approaches \cite{Item-CF} or matrix factorization \cite{MatrixFactor}. However, these shallow models can struggle with complex data. In contrast, embedding-based deep CF—often implemented via a twin-tower model \cite{DSSM, GoogleRetrieval, DAT}—learns user and item embeddings separately and uses their similarity as scores. For large-scale retrieval, approximate nearest neighbor search \cite{fu2021high} provide efficient indexing, making this approach both robust and scalable for personalized retrieval.

\noindent\textbf{Ordinal Logistic Regression (OLR)} models ordinal categories by estimating cumulative probabilities \cite{regression1980}. Enhanced variants (e.g., generalized or partial proportional odds models \cite{partial_proportional_odds_models,PPOM,tutz2022ordinal}) allow category-specific effects, and neural OLR further extends learning capacity \cite{ordinalmodelsforneuralnetworks}. Researchers initially applied OLR to explicit ordered feedback in recommender systems \cite{hu2018collaborative}. Later work adapted OLR for implicit signals, for instance by grouping positive feedback for each item to form ordered labels \cite{parra2011implicit} or designing generalized OLR for sparse targets \cite{faletto2023predicting}. However, both works can only deal with single feedback. No existing methods address the modern large-scale challenge as GNOLR tackles—unifying multiple implicit feedback types to capture global user preference.

\noindent\textbf{Learning to Rank (LTR)} trains models to order items by user preference or relevance, commonly used in search engines and recommender systems. LTR methods generally fall into three categories: (1) Pointwise methods \cite{MCRank} predict a relevance score for each user-item pair independently and potentially overlooks relative relationships among items. (2) Pairwise methods \cite{RankNet,LambdaMART,LambdaRank} compares sample pairs to determine which item in a pair is more relevant, thereby directly modeling relative ranking. (3) Listwise methods \cite{ListNet, ListMLE} optimize the entire list’s ranking order (items displayed in a user session or under a query can be viewed as a list), aiming to place positive items at higher positions in a global scope. 

\noindent\textbf{Multi-Task Recommendation.} Modern recommender systems often aim to optimize multiple objectives (e.g., click, purchases) simultaneously. Early approaches trained separate models per task, combining outputs via fusion layers, but struggled with label inconsistencies and missed task inter-dependencies. Some Multi-task learning (MTL) methods focus on strict causal dependencies like \cite{ESMM,ESCM2,DCMT}, while others focus on architecture innovation like \cite{MMoE,PLE,TAFE,AITM,ResFlow}. Nonetheless, most methods still treat each task independently, missing the progressive nature of user behaviors.
\section{CONCLUSION}
This paper introduces GNOLR, a versatile embedding-based collaborative filtering methods, including a mapping technique to transform multiple implicit feedback to ordinal categories and a novel category-specialized nested encoding framework to model the progression of user engagement into a unified space. Theoretical comparisons highlight GNOLR’s commonality with, and strengths over, prior paradigms. Extensive experiments confirm GNOLR’s efficiency, adaptability, and robustness, outperforming state-of-the-art methods in diverse settings.

\begin{acks}
This paper is partially supported by National Natural Science Foundation of China (NO. U23A20313, 62372471) and The Science Foundation for Distinguished Young Scholars of Hunan Province (NO. 2023JJ10080)
\end{acks}

% \balance
\bibliographystyle{ACM-Reference-Format}
\bibliography{References}

% \clearpage
% \let\balance\relax
% \let\endbalance\relax

\newpage
\appendix
\section{Appendix}

\subsection{Embedding Based Collaborative Filtering}
\begin{figure}[h]
    \centering
    \includegraphics[width=0.6\linewidth]{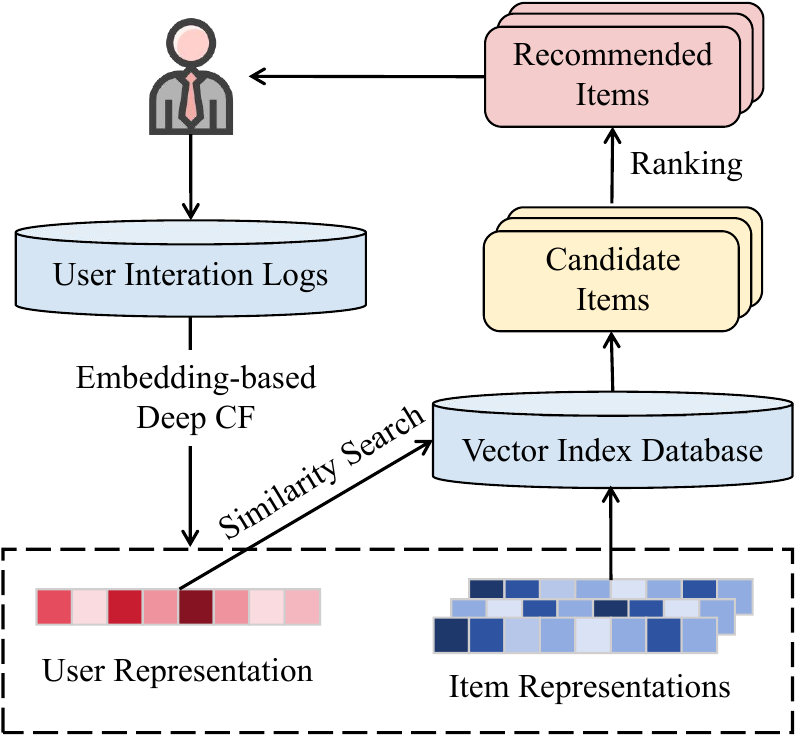}
    \vspace{-2mm}
    \caption{An illustration of real-world recommender system design when applying deep-learning based collaborative filtering to generate relevant candidate items.}
    \Description[An illustration of real-world recommender system design]{This figure depicts the architecture of a real-world recommender system that applies deep learning-based collaborative filtering. It shows how user behavior and item data are processed through deep models to generate relevant candidate items for recommendation. The system is designed to efficiently capture user preferences and provide scalable, personalized suggestions.}
    \label{fig:dlcf}
\end{figure}

Embedding-based collaborative filtering (ECF) is widely deployed in modern recommender systems. Due to the limited expressiveness of linear models, deep learning techniques are widely employed \cite{NCF}. In large-scale, cascade-based pipelines \cite{YoutubeDnn}, ECF often pairs with nearest neighbor search \cite{malkov2018efficient} to produce a relatively small, personalized candidate set for downstream ranking. As illustrated in Figure~\ref{fig:dlcf}, the workflow typically involves offline and online stages. 

\textbf{Offline Stage}: A trained ECF model generates user and item embeddings separately. The item embeddings are then indexed for approximate nearest neighbor search. 

\textbf{Online Stage}: User embeddings are computed on the fly to capture instant preferences, and nearest neighbor search retrieves relevant items from the indexed embedding database. Because of strict latency constraints, the candidate set must be truncated before fine-grained ranking.

This coarse-grained candidate generation imposes two key requirements. \textbf{(1) Independent Embeddings}: User and item embeddings must be generated independently, precluding user–item cross interaction features to avoid train–inference mismatch. \textbf{(2) Euclidean (or Isomorphic) Metric Space}: Indexing methods typically rely on Euclidean distance \cite{malkov2018efficient,fu12fast,fu2021high,dasgupta2011fast,jegou2010product,ge2013optimized}. Cosine similarity is isomorphic to Euclidean distance because vectors are $\ell_2$-normalized, ensuring higher cosine similarity corresponds to a smaller Euclidean distance. 

Prior paradigm (feedback-wise) ECF uses multiple feedback to model user preference. Each feedback type would require its own index, increasing system complexity and leading to suboptimal performance due to separate retrieval and truncation. In addition, a fusion heuristic is needed to merge the results from the multiple independent retrieval process. A typical fusion strategy is a multiplicative formula. For example,
\begin{equation}
    Score_{final}=Score_{task_1}^{\alpha}\cdot Score_{task_2}^{\beta}\cdot Score_{task_3}^{\gamma} \notag
\end{equation}
Such formula may not fully capture the complex relationship between feedback and suffers from precision issues and inefficient parameter searching as the number of tasks grows. This underscores the importance of our proposed approach.

\begin{table}[t]
\centering
\caption{Statistics of all datasets. The number of users and items in the AE series datasets is not disclosed. \#Pos$_1$ and \#Pos$_2$ denotes the counts for click and pay in e-commercial datasets, likes and follows for video datasets.}
\label{tab:data-detail}
\resizebox{\columnwidth}{!}{
\begin{tabular}{ccccccc}
\hline
\hline
Dataset                  & \#User                 & \#Item                 & Split & Total & \#Pos$_1$  & \#Pos$_2$ \\ \hline
\multirow{2}{*}{Ali-CCP} & \multirow{2}{*}{0.24M} & \multirow{2}{*}{0.47M} & train & 41.9M & 1.63M      & 8.4K  \\
                         &                        &                        & test  & 27.2M & 0.99M      & 4.7K  \\ \hline
\multirow{2}{*}{AE-ES}   & \multirow{2}{*}{/}     & \multirow{2}{*}{/}     & train & 22.3M & 0.57M      & 12.9K \\
                         &                        &                        & test  & 9.3M  & 0.27M      & 6.2K  \\ \hline
\multirow{2}{*}{AE-FR}   & \multirow{2}{*}{/}     & \multirow{2}{*}{/}     & train & 18.2M & 0.34M      & 9.1K  \\
                         &                        &                        & test  & 8.8M  & 0.20M      & 5.4K  \\ \hline
\multirow{2}{*}{AE-NL}   & \multirow{2}{*}{/}     & \multirow{2}{*}{/}     & train & 1.8M  & 0.035M     & 1.2K  \\
                         &                        &                        & test  & 5.6M  & 0.14M      & 4.9K  \\ \hline
\multirow{2}{*}{AE-US}   & \multirow{2}{*}{/}     & \multirow{2}{*}{/}     & train & 20M   & 0.29M      & 7.0K  \\
                         &                        &                        & test  & 7.5M  & 0.16M      & 3.9K  \\  \hline
\multirow{2}{*}{KR-Pure} & \multirow{2}{*}{24.9K} & \multirow{2}{*}{6.8K}  & train & 0.72M & 15.8K      & 0.86K \\
                         &                        &                        & test  & 0.2M  & 4K         & 0.28K \\ \hline
\multirow{2}{*}{KR-1K}   & \multirow{2}{*}{1K}    & \multirow{2}{*}{2.4M}  & train & 4.9M  & 94K        & 5.7K  \\
                         &                        &                        & test  & 1.8M  & 35K        & 2.3K  \\ \hline
\multirow{2}{*}{ML-1M}   & \multirow{2}{*}{6K}    & \multirow{2}{*}{3.7K}  & train & 0.7M  & 0.17M      & /     \\
                         &                        &                        & test  & 0.3M  & 0.058M     & /     \\ \hline
\multirow{2}{*}{ML-20M}  & \multirow{2}{*}{138K}  & \multirow{2}{*}{27K}   & train & 13.9M & 3.2M       & /     \\
                         &                        &                        & test  & 6.1M  & 1.3M       & /     \\ \hline
\multirow{2}{*}{RetailR} & \multirow{2}{*}{1.4M}  & \multirow{2}{*}{0.24M} & train & 2.13M & 56K        & 9.8K  \\
                         &                        &                        & test  & 0.63M & 36K        & 12.7K \\ \hline
\hline
\end{tabular}}
\end{table}

\begin{table}[t]
\centering
\caption{Number of features used in each dataset.}
\vspace{-2mm}
\resizebox{\columnwidth}{!}{
\label{tab:feature_usage}
\begin{tabular}{ccccccc}
\hline
\hline
Feature Type & AliCCP & AE & ML-1M & ML-20M & KR & RetailR \\
\hline
User Feature & 9 & 33 & 4 & 1 & 30 & 1 \\
Item Feature & 5 & 47 & 2 & 2 & 63 & 1\\
\hline
\hline
\end{tabular}}
\end{table}

\subsection{Reproducibility}
\subsubsection{Dataset Preprocess}
To comply with the independent embedding requirement, we omit cross features. Specifically, the AliCCP dataset includes 4 cross features, named as 508, 509, 702, 853, which are not used in our experiments. For the rest datasets, there is no such features, thus we use all the features. To maintain consistent learning speed among features and mitigate outlier effects, numerical features are discretized into 50 bins using percentile-based bucketization on the training set. The same bin boundaries are then applied to the test set. Table~\ref{tab:feature_usage} summarizes the quantity of used features for each dataset. 

For the AE series datasets, which originate from a personalized e-commerce search scenario (user search merchandises with a query), each user–query pair is treated as a distinct user to align with the modeling on other datasets.

If official splits are provided, we adopt them directly. Otherwise (e.g., MovieLens), we perform the chronological split from Section~\ref{sec:dataset}. A 10\% subset of the training data serves as the validation set for hyperparameter tuning. A detailed version of the datasets are given in Table \ref{tab:data-detail}.

\subsubsection{Hyperparameters}
We use a batch size of 1024 for pointwise single-task and multi-task training to ensure at least one positive sample per feedback in each batch. For listwise and retrieval experiments, where the model compares positive and negative samples from the same user, we restructure each user’s interaction history into lists, preserving intra-user interactions. To prevent excessively long sequences, user interaction lists exceeding 500 items are randomly split into smaller sublists. We then adopt a batch size of 32 lists, enabling effective listwise and pairwise comparisons within each batch.

All methods employ the Adam optimizer, with hyperparameters determined via grid search for each dataset and method over 200 epoch runs. For all methods, the optimal learning rate is 0.05 (ML-1M), 0.5 (ML-20M), 0.01 (AliCCP), 0.05 (KuaiRand), 5 (AE) and 0.05 (RetailRocket).

\begin{figure}[t]
    \centering
    \includegraphics[width=0.7\linewidth]{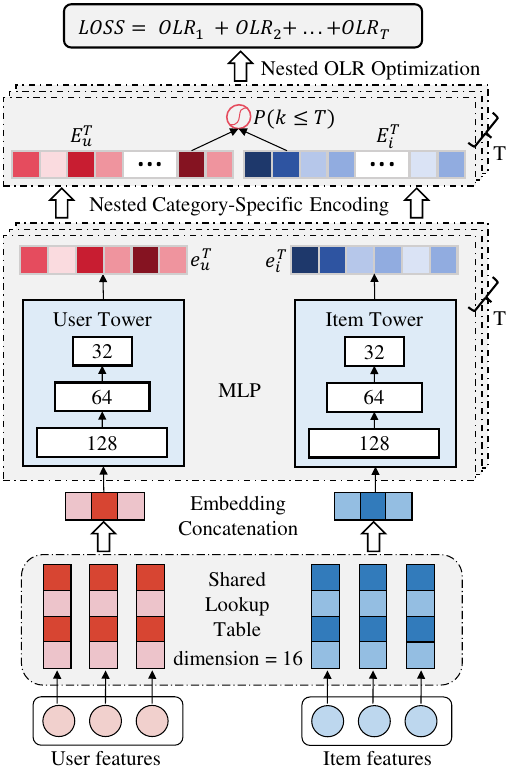}
    \vspace{-2mm}
    \caption{The implementation details of the GNOLR, where the twin tower architecture is consistent across all methods.}
    \Description[GNOLR implementation structure with twin tower architecture]{This figure illustrates the implementation details of the GNOLR model. It adopts a twin tower architecture, where user and item features are encoded separately before interaction. The structure is consistent across all methods being compared, ensuring a fair evaluation of model variants.}
    \vspace{-4mm}
    \label{fig:structure}
\end{figure}

For all methods, we adopt a Twin Tower base architecture with an embedding dimension of 16 and an MLP structure of $\{128,64,32\}$, as illustrated in Figure~\ref{fig:structure}. The embedding lookup tables for each feature are randomly initialized and shared across tasks. We use LeakyReLU activations and no dropout. Our experiments show that these methods exhibit low sensitivity to variations in the network backbone. Thus we use the same backbone for all methods and all datasets, except that we use $\{256,128,64\}$ MLP for naive Neural OLR on multi-task (two tasks) setting to ensure the fairness in parameter quantity. \textbf{The code} is provided in \url{https://github.com/FuCongResearchSquad/GNOLR}

For GNOLR, additional hyperparameters must be specified: $\{a_c\}$ and $\gamma$. We directly set $\{a_c\}$ according to the calculation in Section~\ref{sec:cate-threshold} and perform a grid search to find the optimal $\gamma$. Table~\ref{tab:gnolr_paramenters} summarizes these configurations.

\begin{table}[t]
\centering
\caption{Optimal $a_c$ (Section \ref{sec:cate-threshold}) and $\gamma$ (grid-search) of GNOLR used in different datasets and tasks. The listwise and retrieval tasks use the same configuration as the single-task.}
\label{tab:gnolr_paramenters}
\begin{tabular}{ccccccccccc}
\hline
\hline
\multirow{2}{*}{Dataset} & \multicolumn{2}{c}{Single-Task} & \multicolumn{3}{c}{Multi-Task} \\ \cmidrule(l){2-3} \cmidrule(l){4-6} 
                         & $a_1$         & $\gamma$        & $a_1$   & $a_2$   & $\gamma$   \\ \hline
AliCCP                   & 3.2343        & 7.00            & 3.2343  & 8.5681  &  4.0       \\
AE-ES                    & 3.6003        & 7.00            & 3.6003  & 7.4130  &  1.0       \\
AE-FR                    & 3.8880        & 7.00            & 3.8880  & 7.5351  &  3.0       \\
AE-US                    & 4.0931        & 8.00            & 4.0931  & 7.8353  &  2.0       \\
AE-NL                    & 3.7294        & 2.80            & 3.7294  & 7.0835  &  3.0       \\
KR-Pure                  & 3.8181        & 7.30            & 3.8181  & 6.6975  &  2.8       \\
KR-1K                    & 3.9332        & 2.80            & 3.9332  & 6.7388  &  2.5       \\
ML-1M                    & 1.2295        & 2.57            & /       & /       & /          \\
ML-20M                   & 1.2560        & 3.00            & /       & /       & /          \\
RetailR                  & 3.3682        & 2.00            & 3.3682  & 4.8018  &  2.0       \\ \hline
\hline
\end{tabular}
\end{table}

\begin{table}[t]
\centering
\caption{The AUC performance comparison of various listwise ranking methods across multiple datasets. The best results (statistically significant) are highlighted in bold, while the second-best are underlined. $\lambda$Rank is LambdaRank.}
\label{tab:single-task-listrank-auc}
\resizebox{\columnwidth}{!}{
\begin{tabular}{cccccccccc}
\hline
\hline
Dataset       & RankNet      & $\lambda$Rank   & ListNet      & S2SRank      & SetRank         & JRC          & GNOLR$_L$      \\ \hline
AliCCP        & {\ul 0.5696} & 0.5681          & 0.5577       & 0.5658       & 0.5682          & 0.5009       & \textbf{0.6232} \\
AE-ES         & {\ul 0.7332} & 0.7297          & 0.6403       & 0.7293       & 0.7200          & 0.6323       & \textbf{0.7388} \\
AE-FR         & {\ul 0.7330} & 0.7277          & 0.7298       & 0.7294       & 0.7216          & 0.6428       & \textbf{0.7385} \\
AE-US         & {\ul 0.7046} & 0.6996          & 0.6924       & 0.7009       & 0.6874          & 0.6256       & \textbf{0.7061} \\
AE-NL         & {\ul 0.6923} & 0.6897          & 0.6583       & 0.6915       & 0.6831          & 0.6419       & \textbf{0.7301} \\
KR-Pure       & 0.6396       & 0.6345          & 0.6355       & 0.6408       & 0.6384          & {\ul 0.7835} & \textbf{0.8326} \\
KR-1K         & 0.6321       & 0.5611          & 0.6531       & 0.7047       & 0.5354          & {\ul 0.7964} & \textbf{0.8834} \\
ML-1M         & 0.7477       & 0.7327          & 0.7305       & 0.7443       & 0.7409          & {\ul 0.7937} & \textbf{0.8080} \\
ML-20M        & 0.7138       & 0.7007          & 0.7101       & 0.7187       & 0.7154          & {\ul 0.7801} & \textbf{0.8072} \\
RetailR       & 0.6292       & 0.6238          & 0.5592       & 0.6256       & 0.6136          & {\ul 0.7275} & \textbf{0.7294} \\ \hline
\hline
\end{tabular}}
\end{table}

\subsubsection{Training Stability}
Because $P(\rv{k}=c|\vecrv{x})=P(\rv{k}\le c|\vecrv{x})-P(\rv{k}\le c-1|\vecrv{x})$, the negative log-likelihood requires that this difference remain positive. To prevent numerical instability, we clip these probabilities to $(1e^{-6},\infty)$.

\subsection{Additional Experimental Results}
\subsubsection{Listwise GNOLR} GNOLR is inherently a pointwise ranking and retrieval method, which does not explicitly enforce user-specific dependencies. However, it can be combined with a listwise loss to enhance personalization, as demonstrated in our experiments. Specifically, in the OLR framework, $\gamma\mathcal{K}(\vecmy{E}_u^c,\vecmy{E}_i^c)$ serves as the model’s logit. To incorporate listwise learning, we reuse this logit in a ListNet-style loss:
\begin{equation}
    Loss_{\text{ListNet}} = -\sum_{l=1}^L\sum_{x^+\in\mathcal{P}}\frac{e^{\gamma\mathcal{K}(\vecmy{E}_u^+,\vecmy{E}_i^+)}}{\sum_{x\in \mathcal{P}\cup\mathcal{N}} e^{\gamma\mathcal{K}(\vecmy{E}_u,\vecmy{E}_i)}},
\end{equation}
where $L$ is the number of lists, $\mathcal{P}$ denotes the set of positive samples within one list, and $\mathcal{N}$ denotes the set of negative samples  within one list. The final loss function combines the GNOLR loss and the listwise loss:
\begin{equation}
    Loss=Loss_{\text{GNOLR}}+Loss_{\text{ListNet}}
\end{equation}

\begin{table}[t]
\centering
\caption{The positive sample weight for each task and dataset, where the negative sample weight is fixed as 1. For example, we set weight=10 for positive clicked samples and weight=100 for positive purchased samples. BCE and JRC are only evaluated under the single-task setting.}
\label{tab:pos_weight}
\resizebox{\columnwidth}{!}{
\begin{tabular}{ccccccc}
\hline
\hline
Dataset                     & AliCCP       & AE           & KR-Pure      & KR-1K        & ML      & RetailR      \\ \hline
BCE                         & {[}10{]}     & {[}10{]}     & {[}10{]}     & {[}10{]}     & {[}2{]} & {[}10{]}     \\
JRC                         & {[}10{]}     & {[}10{]}     & {[}10{]}     & {[}10{]}     & {[}2{]} & {[}10{]}     \\
NSB                         & {[}10,100{]} & {[}100,10{]} & {[}100,1{]}  & {[}50,100{]} & /       & {[}10,100{]} \\
ESMM                        & {[}10,50{]}  & {[}10,50{]}  & {[}100,10{]} & {[}100,50{]} & /       & {[}10,50{]}  \\
ESCM\textsuperscript{2}-IPS & {[}10,50{]}  & {[}15,50{]}  & {[}100,1{]}  & {[}50,100{]} & /       & {[}10,50{]}  \\
ESCM\textsuperscript{2}-DR  & {[}10,50{]}  & {[}50,100{]} & {[}100,1{]}  & {[}50,100{]} & /       & {[}10,50{]}  \\
DCMT                        & {[}10,50{]}  & {[}10,50{]}  & {[}100,1{]}  & {[}50,100{]} & /       & {[}10,50{]}  \\
NISE                        & {[}10,50{]}  & {[}10,50{]}  & {[}50,1{]}   & {[}50,100{]} & /       & {[}5,50{]}   \\
TAFE                        & {[}10,50{]}  & {[}10,50{]}  & {[}50,20{]}  & {[}50,100{]} & /       & {[}10,50{]}  \\ \hline
\hline
\end{tabular}}
\end{table}

\begin{table}[t]
\centering
\caption{Performance comparison with multiple implicit feedback types on KR-1K. Single refers to single task prediction for each feedback as baseline.}
\label{tab:multi-feedback}
\begin{tabular}{ccccc}
\hline
\hline
       & Click           & Like            & Follow          & Forward         \\ \hline
Single & {\ul 0.7061}    & {\ul 0.8934}    & {\ul 0.7532}    & {\ul 0.7495}    \\
NSB    & 0.7024          & 0.8885          & 0.7197          & 0.5991          \\
ESMM   & 0.6839          & 0.8916          & 0.7354          & 0.5967          \\
TAFE   & 0.6540          & 0.8576          & 0.7326          & 0.7339          \\
GNOLR  & \textbf{0.7097} & \textbf{0.9101} & \textbf{0.9023} & \textbf{0.7947} \\ \hline
\hline
\end{tabular}
\end{table}

\begin{table}[t]
\centering
\caption{Performance comparison between Manual $\{a_c\}$ and learned on KR-1K.}
\label{tab:learnd-a}
\begin{tabular}{cccc}
\hline
\hline
(auc/$\{a_c\}$) & Manual              & Learned    & Learned w/ target \\ \hline
Like            & \textbf{0.9120}/3.9 & 0.9098/3.0 & 0.9114/2.6        \\
Follow          & \textbf{0.8204}/6.7 & 0.7233/3.7 & 0.7566/6.3        \\ \hline
\end{tabular}
\end{table}

\subsubsection{Listwise AUC} Table~\ref{tab:single-task-listrank-auc} shows the AUC results for the listwise experiment, complementing the GAUC results in Table~\ref{tab:single-task-listrank}. Pairwise, setwise, and listwise baselines struggle to balance AUC (calibration) and GAUC (personalization). By contrast, GNOLR$_L$ increases GAUC without degrading AUC, demonstrating a more effective balance.

\subsubsection{Re-weighting in Multi-Task} Table~\ref{tab:noPosWeight} reports the multi-task AUC without sample re-weighting, complementing the single-task results in Table~\ref{tab:single-task-pointwise} and~\ref{tab:multi-task}. While sample re-weighting improves cross-entropy-based baselines' performance on imbalanced data, it introduces two main drawbacks. 

\textbf{(1)} The optimal positive weights vary widely by dataset and task (presented in Table~\ref{tab:pos_weight}), making Pareto optimization difficult—especially as tasks grow and user distributions evolve in real-world scenarios. GNOLR avoids these complications, achieving its best performance without re-weighting.

\begin{figure}[t]
    \centering
    \includegraphics[width=1\linewidth]{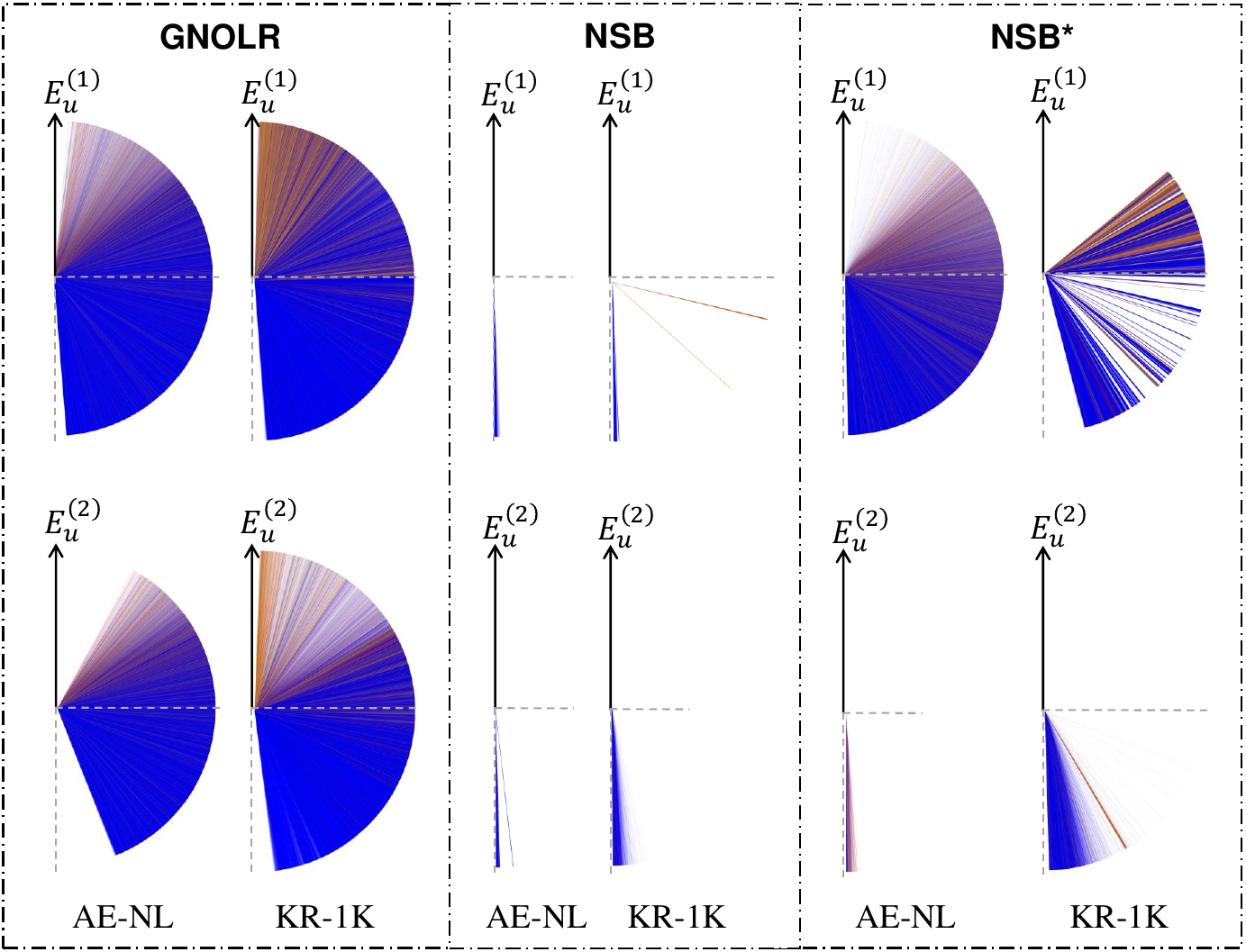}
    \vspace{-2mm}
    \caption{Visualization of the angular distribution between user and item embeddings on KR-1K and AE-NL datasets for different methods. $\vecrv{E}_u^{(1)}$ is the sub-embedding of GNOLR, and $\vecrv{E}_u^{(2)}$ is the unified embedding of GNOLR. For NSB and NSB*, $\vecrv{E}_u^{(1)}$ is the embedding of the denser task, and $\vecrv{E}_u^{(2)}$ is the embedding of the sparse task. NSB* is the sample re-weighted version of NSB.}
    \Description[Angular distribution visualization of user and item embeddings]{This figure visualizes the angular distributions between user and item embeddings on the KR-1K and AE-NL datasets for multiple methods. For GNOLR, \(\vecrv{E}_u^{(1)}\) represents the sub-embedding, and \(\vecrv{E}_u^{(2)}\) the unified embedding. For NSB and NSB*, \(\vecrv{E}_u^{(1)}\) corresponds to the denser task embedding, while \(\vecrv{E}_u^{(2)}\) corresponds to the sparse task embedding. NSB* denotes the sample re-weighted variant of NSB. The figure highlights differences in embedding alignment across methods and datasets.}
    \label{fig:appendix-emb-pic}
\end{figure}

\textbf{(2)} Re-weighting does not resolve the disjoint embedding space inherent in task-independent modeling. Figure~\ref{fig:appendix-emb-pic} (complementing Figure~\ref{fig:case-study}) visualizes all item embeddings' directions when all user embeddings are fixed to an upward vector. 

Without re-weighting, baselines often push item embeddings near $180^{\circ}$ from the user vector—unusable for nearest neighbor search. Even with optimal re-weighting, dense tasks (e.g., CTR) spread items more favorably (< $45^{\circ}$ from the user), while items in sparse tasks (e.g., conversion) still cluster near $180^{\circ}$. Especially, on the KR-1K dataset, multiple disjoint clusters emerge. In contrast, GNOLR produces a continuous and smooth sector shape, placing positive items closer and negative items farther, thus offering a superior unified embedding space. 

Notably, GNOLR’s sub-embeddings also establish favorable spatial proximity between users and items, reflecting strong alignment with user engagement. The shrinkage of unified embedding distribution compared with the sub-embedding may be due to the impact of reshaping factor $\gamma$.

\begin{figure}[t]
    \centering
    \begin{subfigure}[t]{0.49\columnwidth}
        \centering
        \includegraphics[width=\linewidth]{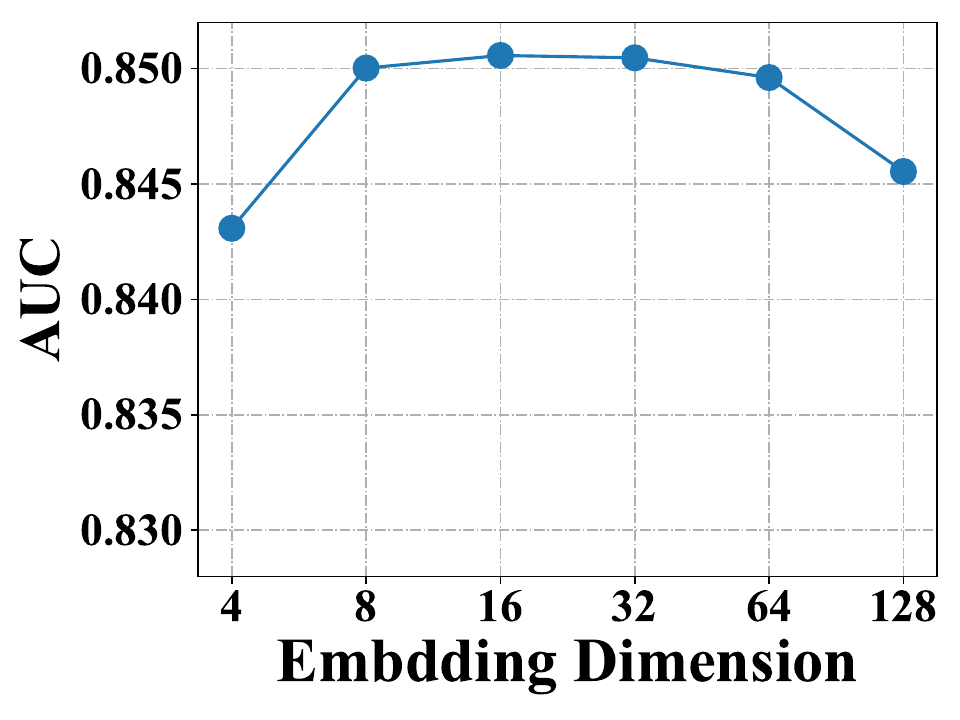}
    \end{subfigure}
    \hfill
    \begin{subfigure}[t]{0.49\columnwidth}
        \centering
        \includegraphics[width=\linewidth]{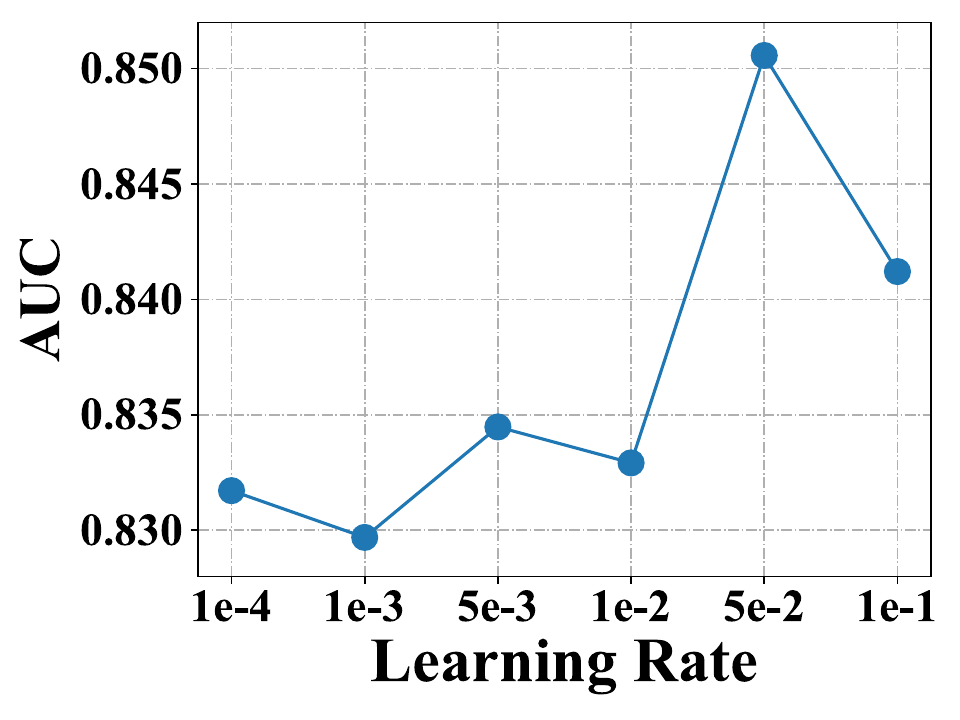}
    \end{subfigure}
    \vspace{-4mm}
    \caption{Parameter sensitivity experiment results on KR-Pure. From left to right: Embdding Dimension, Learning Rate.}
    \Description[Parameter sensitivity results for embedding dimension and learning rate]{This figure presents parameter sensitivity experiments for GNOLR. The left subfigure shows the impact of varying the embedding dimension on model performance, while the right subfigure displays the effect of different learning rates. These results help identify optimal hyperparameter settings for improved model accuracy.}
    \label{fig:hyper-basic}
\end{figure}

\begin{table*}[t]
\centering
\caption{Performance comparison with different backbone architectures on KR-1K.}
\label{tab:backbone_results}
\begin{tabular}{ccccccccccc}
\hline
\hline
Backbone              & Task   & NSB    & ESMM   & ESCM-IPS     & ESCM-DR & DCMT   & NISE   & TAFE   & Neural OLR   & GNOLR           \\ \hline
\multirow{2}{*}{MMoE} & Like   & 0.8867 & 0.8861 & 0.8960       & 0.8975  & 0.9017 & 0.8885 & 0.8915 & {\ul 0.9105} & \textbf{0.9115} \\
                      & Follow & 0.7176 & 0.7164 & {\ul 0.7952} & 0.7783  & 0.7837 & 0.6799 & 0.7204 & 0.6966       & \textbf{0.8188} \\ \hline
\multirow{2}{*}{MMFI} & Like   & 0.8969 & 0.8994 & 0.8988       & 0.8861  & 0.8935 & 0.9020 & 0.9048 & {\ul 0.9108} & \textbf{0.9127} \\
                      & Follow & 0.7432 & 0.7081 & {\ul 0.7950} & 0.7932  & 0.7933 & 0.6870 & 0.7545 & 0.6930       & \textbf{0.8141} \\ \hline
\hline
\end{tabular}
\end{table*}

\begin{table*}[t]
\centering
\caption{Multi-Task Ranking Results (AUC) w/o Pos Sample Weights.}
\label{tab:noPosWeight}
\begin{tabular}{cccccccccc}
\hline
\hline
Task                    & Dataset & NSB          & ESMM         & ESCM\textsuperscript{2}-IPS    & ESCM\textsuperscript{2}-DR     & DCMT         & NISE         & TAFE         & GNOLR           \\ \hline
\multirow{5}{*}{CTR}    & AliCCP  & 0.4994       & 0.4999       & 0.4992       & 0.5004       & 0.5771       & {\ul 0.5919} & 0.5810       & \textbf{0.6153} \\
                        & AE-ES   & 0.5000       & 0.4999       & 0.4999       & 0.4998       & 0.5001       & 0.5003       & {\ul 0.5004}       & \textbf{0.7372} \\
                        & AE-FR   & 0.5006       & 0.5061       & 0.5046       & 0.4998       & {\ul 0.5132} & 0.5006       & 0.5003       & \textbf{0.7370} \\
                        & AE-US   & 0.5003       & 0.5007       & 0.5025       & 0.5008       & {\ul 0.5092} & 0.4993       & 0.5010       & \textbf{0.6971} \\
                        & AE-NL   & 0.5222       & 0.4983       & 0.5397       & 0.4778       & 0.5216       & {\ul 0.5628} & 0.5378       & \textbf{0.7277} \\ \hline
\multirow{5}{*}{CTCVR}  & AliCCP  & 0.5015       & 0.5004       & 0.5003       & 0.5028       & {\ul 0.5637} & 0.5050       & 0.5308       & \textbf{0.5997} \\
                        & AE-ES   & 0.5034       & 0.5037       & 0.5106       & {\ul 0.7036} & 0.5074       & 0.5030       & 0.5007       & \textbf{0.8827} \\
                        & AE-FR   & {\ul 0.6133} & 0.5353       & 0.5122       & 0.5351       & 0.5114       & 0.5194       & 0.5425       & \textbf{0.8793} \\
                        & AE-US   & 0.5022       & {\ul 0.6012} & 0.5307       & 0.5034       & 0.5522       & 0.5063       & 0.5044       & \textbf{0.8663} \\
                        & AE-NL   & 0.5195       & 0.5216       & 0.5546       & {\ul 0.7022} & 0.5712       & 0.6337       & 0.5900       & \textbf{0.8343} \\ \hline
\multirow{2}{*}{Like}   & KR-Pure & 0.5099       & 0.4989       & {\ul 0.5224} & 0.4977       & 0.4937       & 0.4990       & 0.4985       & \textbf{0.8456} \\
                        & KR-1K   & 0.5689       & {\ul 0.6806} & 0.5720       & 0.5125       & 0.5555       & 0.5540       & 0.6720       & \textbf{0.9087} \\ \hline
\multirow{2}{*}{Follow} & KR-Pure & 0.5334       & 0.5435       & 0.5384       & 0.5365       & {\ul 0.5447} & 0.5378       & 0.5315       & \textbf{0.7161} \\
                        & KR-1K   & 0.5981       & 0.6209       & 0.5996       & 0.5806       & 0.5844       & 0.5937       & {\ul 0.6260} & \textbf{0.8165} \\ \hline
\hline
\end{tabular}
\end{table*}

\subsubsection{Additional Parameter Sensitivity Experiments} Figure~\ref{fig:hyper-basic} shows that GNOLR’s performance remains relatively stable with varying embedding sizes, while an optimal learning rate exists for GNOLR, which is the same for other baselines.

\subsubsection{Experiments with Alternative Backbone Architectures}
To further validate the architectural flexibility of GNOLR, we conduct additional experiments replacing the original backbone (Twin Tower encoders with standard MLPs) with two alternative architectures: MMoE~\cite{MMoE} and MMFI~\cite{MMFI}. All compared methods (including baselines) are re-implemented with these backbones under identical training protocols. As summarized in Table~\ref{tab:backbone_results}, GNOLR consistently outperforms baselines across all backbone architectures, with all methods showing improved performance from advanced structures.

\subsubsection{Extension to Multiple Feedback Types}
While most baselines in the main experiments only model two types of implicit feedback, we further evaluate GNOLR's scalability with \textbf{four implicit feedback types} on the KR-1K dataset. We select adaptably designed baselines (NSB, ESMM, TAFE) and modify their implementations for multi-feedback processing. As shown in Table~\ref{tab:multi-feedback}, GNOLR demonstrates a strong ability to model long-range and progressive implicit user preferences.

\subsubsection{Learning vs. Manual Selection for Ordinal Thresholds $\{a_c\}$}
As mentioned in Section~\ref{sec:cate-threshold}, the thresholds $\{a_c\}$ can be learned jointly with the model. To validate whether our manually selected $\{a_c\}$ is reasonable, we implemented two parameter learning schemes for comparison with the manual selection approach: (1) simply preserving the inherent order constraint of $\{a_c\}$, and (2) additionally imposing a regularization penalty to push the learned $\{a_c\}$ closer to the manual value. The result are shown in Table~\ref{tab:learnd-a}, where we observe that the model can hardly learn optimal $\{a_c\}$ for sparse targets.

\end{document}